\documentclass[aps,prl,twocolumn]{revtex4}
\usepackage{amsfonts}
\usepackage{amsmath}
\usepackage{graphicx}
\usepackage{dsfont}
\usepackage{diagbox}
\usepackage{array}
\usepackage{amssymb}
\usepackage{bbm}
\usepackage{float}
\usepackage{subfigure}
\usepackage{url}
\usepackage{hyperref}

\newcommand{\PreserveBackslash}[1]{\let\temp=\\#1\let\\=\temp}
\newcolumntype{C}[1]{>{\PreserveBackslash\centering}p{#1}}
\newcolumntype{R}[1]{>{\PreserveBackslash\raggedleft}p{#1}}
\newcolumntype{L}[1]{>{\PreserveBackslash\raggedright}p{#1}}

\begin{document}

\title{Tricritical point with fractional supersymmetry from a Fibonacci topological state}
\author{Wen-Tao Xu$^{1}$ and Guang-Ming Zhang$^{1,2}$}
\affiliation{$^{1}$State Key Laboratory of Low-Dimensional Quantum Physics and Department
of Physics, Tsinghua University, Beijing 100084, China\\
$^{2}$Collaborative Innovation Center of Quantum Matter, Beijing 100084,
China.}
\date{\today}

\begin{abstract}
We consider a generic Fibonacci topological wave function on a square
lattice, and the norm of this wave function can be mapped into the partition
function of two-coupled $\phi ^{2}$-state Potts models with $\phi =( \sqrt{5}%
+1)/2$ as the golden ratio. A global phase diagram is thus established to
display non-abelian topological phase transitions. The Fibonacci topological
phase corresponds to an emergent new phase of the two-coupled Potts models,
and continuously change into two non-topological phases separately, which
are dual each other and divided by a first-order phase transition line.
Under the self-duality, the Fibonacci topological state enters into the
first-order transition state at a quantum tricritical point, where two
continuous quantum phase transitions bifurcate. All the topological phase
transitions are driven by condensation of anyonic bosons consisting of
Fibonacci anyon and its conjugate. However, a fractional supersymmetry
is displayed at the quantum tricritical point, characterized by a coset
conformal field theory.
\end{abstract}

\maketitle

\textit{Introduction}. - In recent years theoretical and experimental search
for topological quantum phases of matter with fractionalized anyonic
excitations has attracted considerable attention, because non-abelian
quasiparticles are necessary ingredient for topological quantum computation
\cite{kitaevToricCode2003,freedman2003,RMP_TQC}. The Fibonacci anyons are
the simplest example for universal quantum computation. The Fibonacci
topological phase contains two types of excitations: the trivial one $1$ and
non-abelian anyon $\tau $ with quantum dimension $d_{\tau }=\phi $, where $%
\phi =(\sqrt{5}+1)/2$ is the golden ratio. They satisfy the non-abelian
fusion rule:
\begin{equation}
1\otimes \tau =\tau \otimes 1=\tau, \hspace{0.5cm} \tau \otimes \tau
=1\oplus \tau .
\end{equation}%
The doubled Fibonacci (DFib) phase has two additional anyons: $\bar{\tau}$
with the opposite chirality to $\tau $ and anyon $b=\tau\otimes \bar{\tau}$
as a composite boson. A microscopic lattice model for the DFib topological
phase is given by the Levin-Wen string-net model\cite{levin_string-net_2005}

Since these non-abelian topological phases cannot be described by any local
order parameters, the corresponding phase transitions to their adjacent
non-topological phases are beyond the conventional Landau-Ginzburg-Wilson
paradigm. Previous investigations are devoted to the perturbed Levin-Wen
string-net with a competing interaction\cite%
{J_Vidal_Golden_string_net,Vidal2015}, the wave function deformed by a
string tension\cite{Condensation_driven}, and a reduced one-dimensional
models\cite{GoldenChain,Trebst,gils_topology-driven_2009}. However, because
these models are mostly defined on the trivalent lattice, quantum
self-duality as the important feature of the generic doubled topological
phases\cite{fendley_topological_2008,zhu_gapless_2019} is lacking, i.e., the
invariance of interchanging the vertices and plaquettes in the microscopic
lattice models. So a generic phase diagram of the DFib topological state has
not been established yet.

To incorporate the quantum self-duality into the DFib topological phase, P.
Fendley proposed a quantum net wave function\cite%
{fendley_topological_2008,fendley_fibonacci_2013}. On a square lattice, the
corresponding parent Hamiltonian involving simpler interactions than the
string-net model. In this Letter, based on this quantum net wave function,
we propose a generic DFib topological wave function with two deformed
parameters, and fully develop a wave function approach to the non-abelian
topological phase transitions. The norm of this wave function can be mapped
into the partition function of two-coupled $\phi ^{2}$-state Potts models,
and a generic phase diagram is established. The DFib topological phase
represents an emergent new phase of the two-coupled Potts models, while two
non-topological phases correspond to the low- and high-temperature phases of
the coupled Potts model, respectively, which are dual each other and
separated by a first-order transition line. Along the quantum self-duality
path, a quantum tricritical point exists between the DFib topological phase
and the first-order transition state. Away from the self-duality, we perform
the numerical calculations to determine the phase boundaries via the
transfer operator of the tensor network method\cite{PEPSTO}. The
corresponding conformal field theories (CFT) of the quantum criticality are
fully discussed.

\begin{figure}[tbp]
\centering
\includegraphics[width=8cm,trim=40 65 0 65,clip]{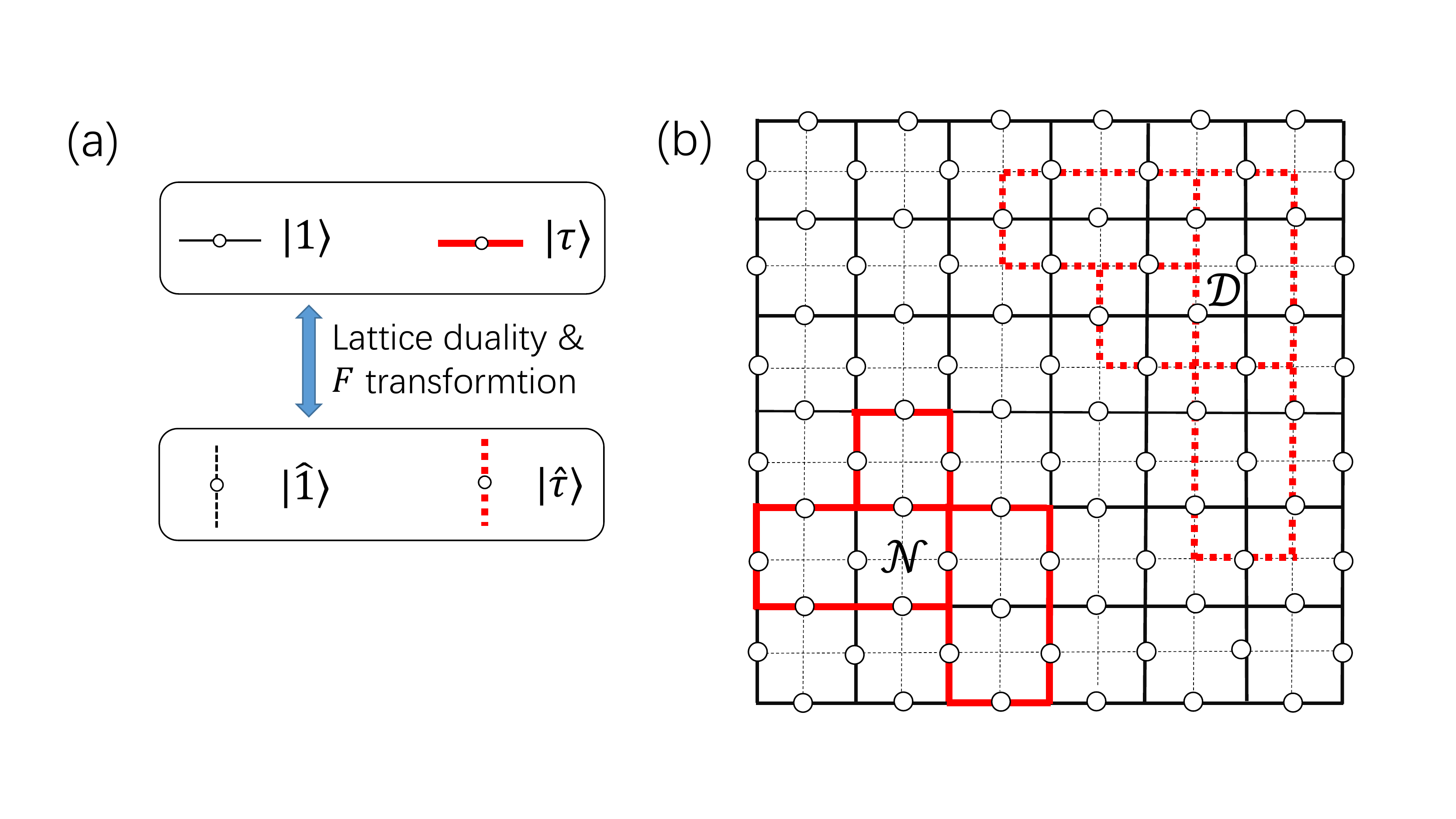}
\caption{(a) Two local orthogonal quantum states $|1\rangle $ and $|\protect%
\tau \rangle $ on an edge of the original lattice, and the transformed
states $|\hat{1}\rangle $ and $|\hat{\protect\tau}\rangle $ after the
quantum duality. (b) The physical degrees of freedom (open circle) locate at
the edges of a square lattice. The full lines are original lattice while the
dash lines are dual lattice. $\mathcal{N}$ is a typical net on the original
lattice and $\mathcal{D}$ is a net on the dual lattice.}
\label{Net}
\end{figure}

\textit{DFib wave function}. - On a square lattice, a quantum net wave
function is given by\cite{fendley_topological_2008,fendley_fibonacci_2013},
\begin{equation}
|\Psi \rangle =\sum_{\mathcal{N}}\phi ^{-L_{\mathcal{N}}/2}\chi _{\hat{%
\mathcal{N}}}\left( \phi ^{2}\right) |\mathcal{N}\rangle ,
\label{Fendley_wavefunction}
\end{equation}%
which is the superposition of nets $\mathcal{N}$. An edge of the lattice is
either empty or occupied by $\tau $ string, yielding two orthogonal quantum
states labelled by $|1\rangle $ and $|\tau \rangle $, see Fig.\ref{Net}(a).
A net is formed by closed $\tau $ strings which are allowed to branch and
cross, as shown in Fig.\ref{Net} (b). The superposition weights are given by
both chromatic polynomial $\chi _{\hat{\mathcal{N}}}(Q)$ and a string
tension $\phi ^{-L_{\mathcal{N}}/2}$, where $L_{\mathcal{N}}$ is the total
length of the $\tau $ strings in the net $\mathcal{N}$. The chromatic
polynomial $\chi _{\hat{\mathcal{N}}}(Q)$ can be understood by treating the
strings $\tau $ in the net $\mathcal{N}$ as the boundaries separating different
regions. The dual net $\hat{\mathcal{N}}$ corresponds to a vertex for each
region and an edge connecting each pair of regions sharing a boundary. For
an integer $Q$, $\chi _{\hat{\mathcal{N}}}(Q)$ is the number of ways of
coloring each region with $Q$ colors such that the neighboring regions are
colored differently. Since $\chi _{\hat{\mathcal{N}}}(Q)$ is a polynomial of
$Q$, we can generalize $Q$ to any non-integers, and the present case $Q=\phi$.

The quantum self-duality of $|\Psi\rangle$ is displayed by rewriting on the
dual lattice in a specified representation, whose basis are orthogonal
quantum states $|\hat{1}\rangle $ and $|\hat{\tau}\rangle $ on an edge of
the dual lattice. As shown in Fig.\ref{Net} (a), they are related to the
local states $|1\rangle$ and $|\tau\rangle $ of the original lattice through
the transformation:
\begin{equation}
F=\frac{1}{\phi }\left(
\begin{array}{cc}
1 & \sqrt{\phi } \\
\sqrt{\phi } & -1%
\end{array}%
\right) .  \label{transformation}
\end{equation}
After the duality transformation, the wave function $|\Psi\rangle$ in Eq. (%
\ref{Fendley_wavefunction}) becomes
\begin{equation}
|\hat{\Psi}\rangle\propto\sum_{\mathcal{D}}\phi ^{-L_{\mathcal{D}}/2}\chi _{%
\hat{\mathcal{D}}}\left( \phi ^{2}\right) |\mathcal{D}\rangle,
\end{equation}
where $\mathcal{D}$ is the net formed by $\hat{\tau}$ strings on dual
lattice, as displayed in Fig. \ref{Net} (b). The $|\hat{\Psi}\rangle$ has
the same form with $|\Psi\rangle$, thus $|\Psi\rangle$ possess the quantum
self-duality\cite{fendley_topological_2008,fendley_fibonacci_2013}.

\textit{Deformed DFib wave function}. - To consider the topological phase
transitions out of the DFib phase, we propose a deformed quantum net wave
function. Analogous to the self-dual toric code model\cite%
{GaugingTNS,zhu_gapless_2019}, the deformed wave function is introduced by
using spin polarized filters in two dual spin directions:
\begin{equation}
|\Psi (h,\hat{h})\rangle =\prod_{\text{edges}}P_{\text{edge}}(h,\hat{h}%
)|\Psi \rangle ,  \label{deformation}
\end{equation}%
where $P_{\text{edge}}(h,\hat{h})=1+h\sigma ^{z}+\hat{h}\hat{\sigma}^{z}$ is
the deformation matrix, $h$ and $\hat{h}$ denote the filter strengths in two
spin directions. $\sigma ^{z}$ is the diagonal Pauli matrix in the $%
|1\rangle $ and $|\tau \rangle $ basis. And $\hat{\sigma}^{z}=F\sigma
^{z}F^{-1}$, which is diagonal Pauli matrix in the $|\hat{1}\rangle $ and $|%
\hat{\tau}\rangle $ basis. It is obvious that the quantum duality transforms
$|\Psi (h,\hat{h})\rangle $ into $|\Psi (\hat{h},h)\rangle $, displaying the
quantum self-duality when $h=\hat{h}$.

The squared wave function weights can be interpreted as the Boltzmann
weights, providing a plasma analogy between topological ground states and
the statistical model. So the equal-time correlators of local quantum
operators can be mapped to the correlators of different local operators of
statistical mechanics model. And the quantum phase transitions are captured
by the phase transitions in the statistical model. In the net basis, the
norm of the deformed wave function is expressed as
\begin{equation}
\mathcal{Z}=\sum_{\mathcal{N},\mathcal{N}^{\prime }}\chi _{\hat{\mathcal{N}}%
}\left( \phi ^{2}\right) \chi _{\hat{\mathcal{N}}^{\prime }}\left( \phi
^{2}\right) \prod_{\text{edge}}W_{n_{e},n_{e}^{\prime }},  \label{norm}
\end{equation}%
where $\mathcal{N}$ and $\mathcal{N}^{\prime }$ denote the nets in the bra
and ket layers, and $n_{e}=1$ or $\tau $ when the edge $e$ of the net $%
\mathcal{N}$ is empty or occupied by the $\tau $ string. While the chromatic
polynomials determine the non-local part of the Boltzmann weights, the local
part is given by the matrix
\begin{equation}
W=\left(
\begin{array}{cc}
Q_{11} & Q_{1\tau }/\sqrt{\phi } \\
Q_{\tau 1}/\sqrt{\phi } & Q_{\tau \tau }/\phi%
\end{array}%
\right) ,
\end{equation}%
where $Q_{11}$, $Q_{1\tau }$, $Q_{\tau 1}$ and $Q_{\tau \tau }$ are elements
of the matrix $P^{2}$. It is known that the partition function of a
ferromagnetic $Q$-state Potts model can be expanded as $\mathcal{Z}_{\text{%
Potts}}(K,Q)=\sum_{\mathcal{N}}e^{-KL_{\mathcal{N}}}\chi _{\hat{\mathcal{N}}%
}(Q)$ with the temperature $1/K$. So the partition function (\ref{norm})
describes two-coupled ferromagnetic $\phi ^{2}$-state Potts models, where
the local Boltzmann weight matrix $W$ includes the terms $e^{-KL_{\mathcal{N}%
}}$ and $e^{-KL_{\mathcal{N^{\prime }}}}$ for each $\phi ^{2}$-state Potts
model and coupling between two Potts models.

In the parameter space of ($h,\hat{h}$), two special lines can be found from
the local Boltzmann weight matrix. When the coupling between two Potts
models vanishes, we have two decoupled $\phi ^{2}$-state Potts models, and
the local part of Boltzmann weight $W_{n_{e},n_{e}^{\prime }}$ becomes a
product of two local Boltzmann weights of the $\phi ^{2}$-state Potts model,
\begin{equation}
W\propto \left(
\begin{array}{cc}
1 & e^{-K} \\
e^{-K} & e^{-2K}%
\end{array}%
\right) .
\end{equation}%
Such a condition is equal to $\det P=0$, from which an ellipse equation is
derived
\begin{equation}
h^{2}-2\left( 2\phi -3\right) h\hat{h}+\hat{h}^{2}=1,
\end{equation}%
determining the outer boundary of the phase diagram of the coupled Potts
model. According to the above equation, the critical point $D$ of each $\phi
^{2}$-state Potts model can be determined to $h=\hat{h}=\phi /2\simeq 0.809$%
. In contrast, as the coupling between two $\phi ^{2}$-state Potts models is
strong enough, the coupled model becomes a single $\phi ^{4}$-state Potts
model with the local Boltzmann weight
\begin{equation}
W\propto \left(
\begin{array}{cc}
1 & e^{-K} \\
e^{-K} & e^{-K}%
\end{array}%
\right) ,
\end{equation}%
and another ellipse curve equation can be deduced
\begin{equation}
(h-1)^{2}+(\hat{h}-1)^{2}-2h\hat{h}\frac{\phi -1}{\phi +1}=1.
\end{equation}%
Due to $\phi ^{4}>4$, this $\phi ^{4}$-state Potts
model can only have a first-order phase transition between the low- and
high-temperature gapped phases\cite{wu_potts_1982}, and the phase transition
point $S$ is also determined to $h=\hat{h}=\frac{\phi }{2}\left( \phi -\sqrt{%
\phi }\right) \simeq 0.280$.

More interestingly, along the self-dual line, a new critical point $C$ can
be found at $h=\hat{h}\simeq 0.197$ by considering the level-rank duality
and the $SO(4)_3$ Birman-Murakami-Wenzl algebra for the transfer matrix of
the coupled Potts model in the loop representation\cite{fendley_critical_2008}.
So this quantum critical point divides the self-dual line into two parts: the
gapped state ($0\leqslant h<0.197$) corresponding to the DFib topological state
and another state ($0.197<h<0.809$) corresponding to the non-topological state.
Both the perturbed analysis\cite{dotsenko_coupled_1999} and the first-order
nature of the $\phi ^{4}$-state Potts model suggest that the non-topological
phase has a small gap. In fact, the ending points of this line ($0.197<h<0.809$)
are two distinct critical points, and the non-topological gapped state represents
a first-order phase transition line.

\begin{figure}[tbp]
\centering
\includegraphics[width=8cm, trim=150 250 100 100, clip]{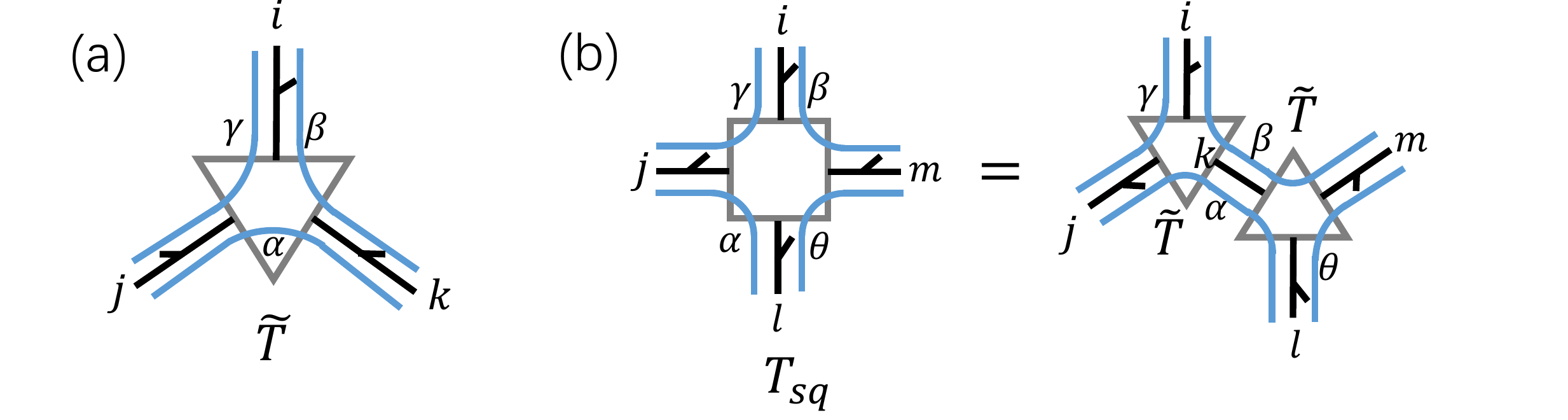}
\caption{(a) The triplet-line local tensor $\tilde{T}_{\protect\alpha
\protect\beta \protect\gamma }^{ijk}$ of the tensor network state for the
string-net model. The legs of the tensor connected by lines (labelled by
only one index for simplicity) are locked by the delta function, and the
legs pointed outside of the paper are physical indices. (b) Contracting two
tensors $\tilde{T}$ along one bond direction of the honeycomb lattice yields
the tensor $T_{sq}$ for the quantum net wave function on a square lattice.}
\label{squrae_tensor}
\end{figure}

\textit{Global phase diagram}. - To establish the global phase diagram, we
have to perform numerical calculations with the deformed DFib wave function.
For simplicity, we will focus on the parameter regime: $h,\hat{h}\geqslant 0$%
. By introducing auxiliary degrees of freedom, the non-local chromatic
polynomials are decomposed into a local structure, yielding a triplet-line
tensor network state. Starting from the local tensor $T_{\alpha \beta \gamma
}^{ijk}$ for the DFib string-net wave function\cite%
{GuTensorNetwork,buerschaper_explicit_2009}, we first define the local
tensor $\tilde{T}_{\alpha \beta \gamma }^{ijk}=\phi ^{-\frac{3}{4}\delta
_{ij}\delta _{jk}\delta _{k\tau }}T_{\alpha \beta \gamma }^{ijk}$. Then, as
shown in Fig. \ref{squrae_tensor}, the local tensor $T_{sq}$ of the quantum
net on the square lattice can be obtained by contracting two triple-line
tensors $\tilde{T}$ along one bond direction of the honeycomb lattice. The
detailed derivation is included in Supplementary Material. Since the
deformation operator $P$ acts on the on-site degrees of freedom, the
deformed local tensor $T_{sq}(h,\hat{h})$ is obtained by simply acting $%
\sqrt{P}$ on the physical indices. The deformed wave function in the tenor
network representation is obtained by contracting all virtual indices $%
\alpha ,\beta ,\gamma $ and $\theta $ of the local tensors
\begin{equation}
|\Psi (h,\hat{h})\rangle =\sum_{\{ij\cdots \}}\text{tTr}\left[ \underset{%
\text{vertex}}{\bigotimes }\phi ^{-\frac{i+j+l+m}{4}}T_{\text{sq}}(h,\hat{h})%
\right] |ij\cdots \rangle .
\end{equation}%
where "tTr" denotes the tensor contraction and $\phi ^{-(i+j+l+m)/4}$ comes
from the string tension of the quantum net $|\Psi \rangle $.

\begin{figure}[tbp]
\centering
\includegraphics[width=8cm]{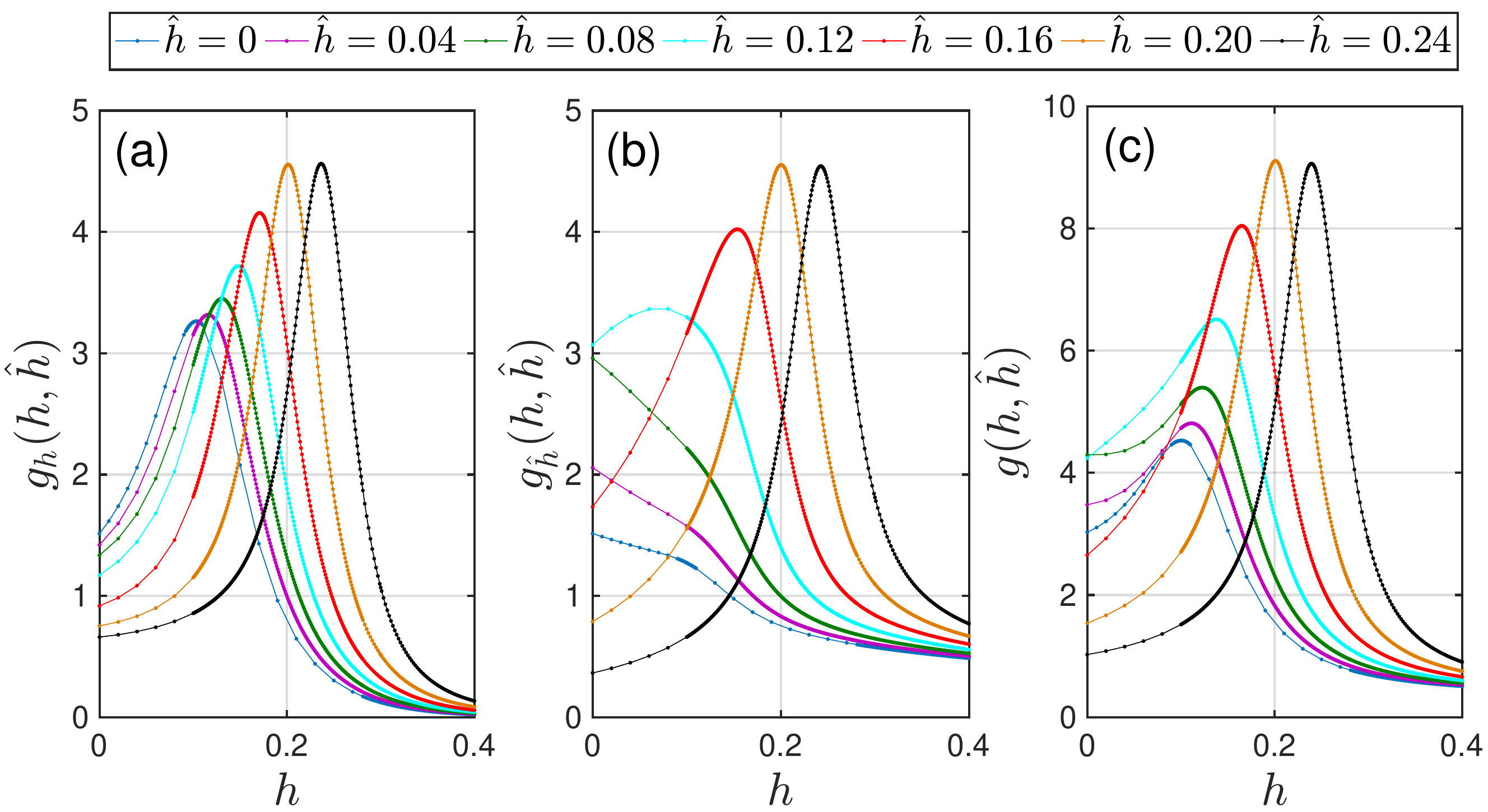}
\caption{The fidelity metric calculated from the wave functions with $%
N=L_yL_x=4\times10$ sites and $\protect\delta h=\protect\delta \hat{h}$%
=0.005, where $\hat{h}$ is fixed and $h$ is varied. (a) One diagonal element
$g_h(h,\hat{h})$. (b) The other diagonal element $g_{\hat{h}}(h,\hat{h})$.
(c) The trace of the fidelity metric $g(h,\hat{h})=g_h(h,\hat{h})+g_{\hat{h}%
}(h,\hat{h})$.}
\label{fidelity}
\end{figure}

Due to lack of local order parameters in the non-abelian topological phase
transitions, we have to determine the phase boundaries by employing the
quantum fidelity\cite{fedility_2007} in the numerical calculation with the
tensor network wave function. The diagonal elements of the quantum fidelity
metric are expressed as
\begin{eqnarray}
g_{h}(h,\hat{h}) &=&-\frac{1}{\left( \delta h\right) {}^{2}N}\log| \langle
\Psi (h,\hat{h})|\Psi (h+\delta h,\hat{h})\rangle| ,  \notag \\
g_{\hat{h}}(h,\hat{h}) &=&-\frac{1}{(\delta \hat{h})^{2}N}\log |\langle \Psi
(h,\hat{h})|\Psi (h,\hat{h}+\delta \hat{h})\rangle| ,  \notag
\end{eqnarray}%
where $N$ is the total number of lattice sites and $\delta h=\delta \hat{h}%
=5\times 10^{-3}$ is chosen in the calculations. Note that all wave
functions in the above expressions are normalized. The quantum duality
implies $g_{h}(h,\hat{h})=g_{\hat{h}}(\hat{h},h)$. In Fig.\ref{fidelity},
for a given value of $\hat{h}$ but different $h$, we show the diagonal
elements of the fidelity metric and its trace value $g_{h}(h,\hat{h})+g_{%
\hat{h}}(h,\hat{h})$, where the peak positions of the curves correspond to
the quantum phase transition points\cite{GaugingTNS}.

Combining the known analytical results on the self-dual line and numerical
results via the quantum fidelity calculations away from the self-duality, we
can establish a global phase diagram, as shown in the Fig. \ref%
{fedility_peak_plus}. Two non-topological gapped phases corresponding to the
low-temperature and high-temperature phases of the Potts model are separated
by a first-order phase transition line $CD$. However, the DFib topological
phase enclosed by two topological phase transition lines $AC$ and $BC$
represents a new phase of the two-coupled Potts model. Along the self-dual
line, the first-order transition line terminates at the critical point $D$.
More importantly, the first-order transition line bifurcates at the quantum
critical point $C$ into two separate topological phase transition
lines $BC$ and $AC$. So this quantum critical point $C$ is a quantum tricritical
point in the phase diagram. The points $A$ and $B$ on the axes are
numerically determined to ($\hat{h}\simeq0.099,h=0$) and ($\hat{h}%
=0,h\simeq0.099$).
\begin{figure}[tbp]
\centering
\includegraphics[width=7cm, trim=10 80 350
100,clip]{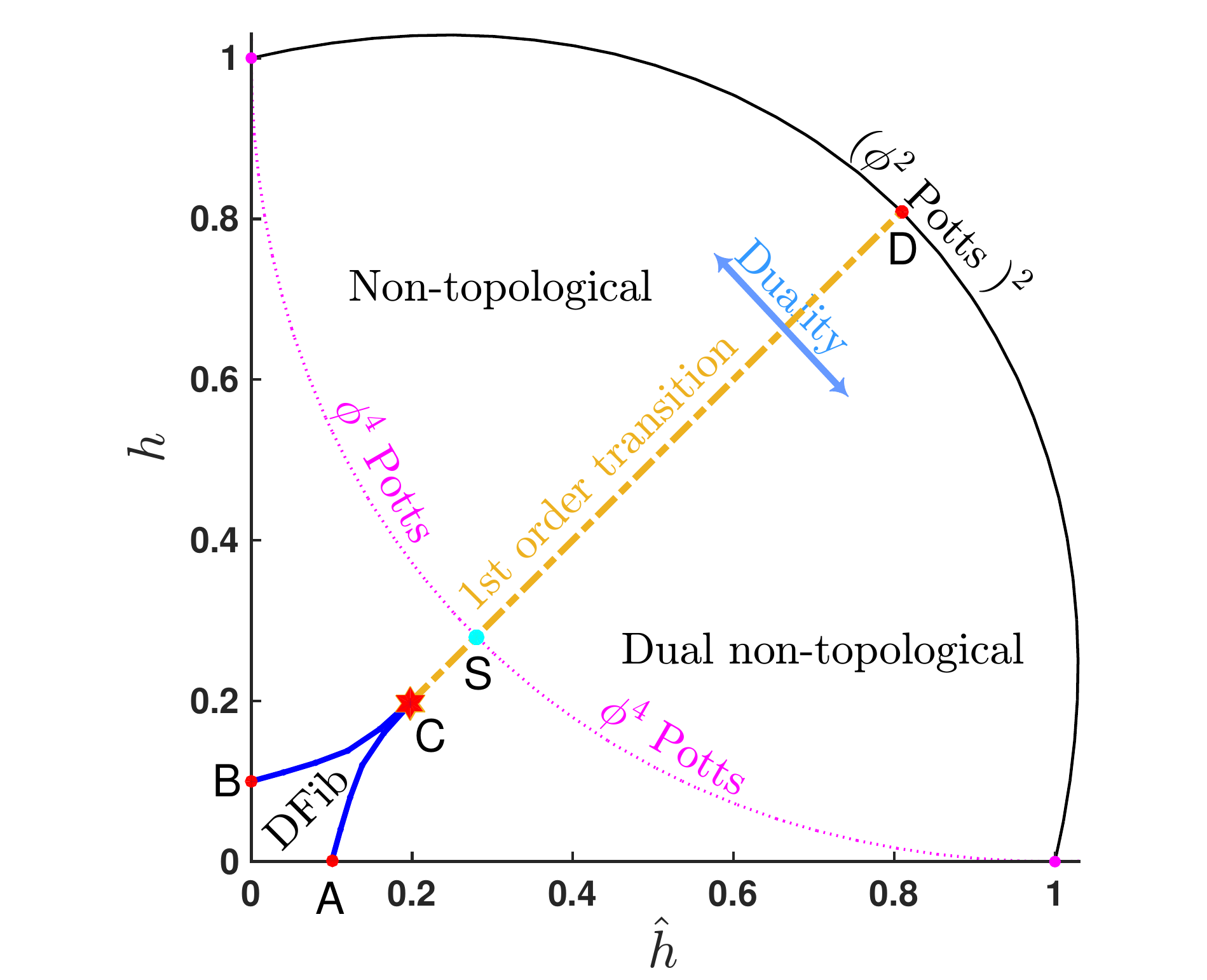}
\caption{The global phase diagram of a generic DFib topological phase. The
second-order phase transition lines $AC$ and $BC$ are determined by the
trace of quantum fidelity metric. Two non-topological phases correspond to
the low- and high-temperature phases of the coupled Potts models, while the
DFib topological phase is an emergent phase of the coupled Potts model. The
point $S$ exactly sits on the first-order transition line, and the point $C$
is the tricritical point. The point $D$ is the critical point of two
decoupled $\protect\phi ^{2}$-state Potts models, and each has a central
charge $c=7/10$}
\label{fedility_peak_plus}
\end{figure}

\textit{Quantum Criticality}. - Along the $h$ axis, the deformation matrix $P
$ is diagonal, the phase transition is equivalent to the DFib topological
phase transition driven by tuning a string tension. The string-net wave
function deformed by a string tension, whose norm is exactly mapped to the $%
(\phi +2)$-state Potts model\cite%
{fidkowski_string_2009,fendley_topological_2008}, has been studied\cite%
{Condensation_driven}. However, due to the difference between the quantum
net wave function $|\Psi\rangle$ on the square lattice and the string-net
one on the honeycomb lattice, the wave function norm of the quantum net can
not be exactly mapped to the partition function of the $(\phi +2)$-state
Potts model. Nevertheless, it is argued that the difference is an irrelevant
perturbation\cite{fendley_topological_2008} so that the phase transition of
the quantum net wave function still belongs to the same universality class%
\cite{Fendley_2005}. Actually our numerical calculations confirm this
conclusion (see Supplementary Material). So the corresponding critical point
is described by the seventh unitary minimal CFT with the central charge $%
c=14/15$. This continuous phase transition is driven by the condensation of $%
b$ anyons consisting of the Fibonacci anyon and its conjugate\cite%
{Condensation_driven}.

Since the quantum critical point $B$ is a conformal quantum critical point%
\cite{ardonne}, all equal-time correlation functions are described by the
two dimensional CFT. From the quantum duality, the quantum critical point $A$
on the dual axis $\hat{h}$ has the same property. The validity of the
universality hypothesis allows us to further argue that the topological
phase transitions across both the $AC$ and $BC$ lines are effectively
characterized by the same critical field theory as the $(\phi +2) $-state
Potts model, as there is no other mechanism for the topological quantum
phase transition. We notice that a similar situation occurs in the
topological phase transitions of the deformed toric-code wave function with
duality\cite{zhu_gapless_2019}.

Moreover, it is known that the level-rank duality of the coupled Potts model
in the loop representation not only determines the position of a tricritical
point but also gives rise to the underlying critical field theory\cite%
{fendley_critical_2008}. Namely it is the coset CFT\cite%
{goodard_virasoro_1985} of $\frac{SU(2)_{3}\times SU(2)_{3}}{SU(2)_{6}}$
with a central charge $c=27/20$, which is regarded as the sum of a bosonic
field with a given background screening charge located at infinity and a $%
\mathbb{Z}_{3}$ parafermionic field\cite%
{ravanini_infinite_1988,Bagger_Virasoro_1988}. This quantum tricritical
point $C$ is also a conformal quantum tricritical point, and the scaling
dimensions of the primary fields together with our numerical results are listed
in Tab.\ref{27/20CFT}.
\begin{table}[tbp]
\caption{Scaling dimensions $\mathfrak{h}_{p,q}+\bar{\mathfrak{h}}_{p,q}$ of
the primary fields in the coset CFT for the quantum
tricritical point. The third columns gives the numerical values extracted
from the transfer operator with a circumference $L_{y}=6$, whose eigenvalue
spectrum is rescaled such that the scaling dimension of the first descendent
for the lowest primary field is unity. The fourth column lists the quantum
numbers of the matrix product operator. The data with the absolute errors
larger than 10\% are indicated by stars, and the last three primary fields
have some difficulties to be determined.}
\label{27/20CFT}\renewcommand\arraystretch{1.2}
\begin{tabular}{C{1cm}C{1.5cm}C{2cm}C{1.5cm}C{2cm}}
\toprule
 $(p,q)$ & $\mathfrak{h}_{p,q}+\bar{\mathfrak{h}}_{p,q}$ & Numerical result & quantum number& Absolute error\\
\colrule
 $(1,1)$ & $0$ &  $0$ &$\phi^2$ &$0$\\

 $(2,3)$ & $1/10$ & $0.10692317$&$-1$&$0.0069$\\

 $(1,2)$ & $9/80$  & $0.11235947$&$\phi^{-2}$&$0.0253$\\

 $(2,2)$ & $9/80$  & $0.11235947$&$\phi^{-2}$&$0.0253$ \\

 $(2,4)$  & $13/80$ & $0.16501496$&$-1$& $0.0025$\\

 $(1,3)$ & $3/10$ &   $0.30662991$&$-1$ & $0.0066$\\

 $(2,5)$ & $3/10$ &   $0.30662991$&$-1$& $0.0066$ \\

 $(1,4)$ & $9/16$ & $0.55837990$&$\phi^{-2}$&$0.0041$    \\

  $(3,7)$ & $3/5$ & $0.67849389$&$\phi^{-2}$&$0.0785$  \\

 $(2,6)$  & $73/80$  & $0.74175242^*$&$\phi^{2}$&$0.1707$  \\

 $(1,5)$ & $13/10$ &   $0.97123493^*$& $-1$&  $0.3288$ \\

 $(2,7)$ & $8/5$&$-$&$-$ &$-$  \\
  $(1,6)$ & $169/80$ &$-$&$-$ &$-$ \\
   $(1,7)$ & $3$ &$-$& $-$&$-$ \\
\botrule
\end{tabular}
\end{table}

Surprisingly, there emerges a fractional superconformal symmetry at the
quantum tricritical point $C$. Generally, the coset CFT of $\frac{%
SU(2)_{M}\times SU(2)_{N}}{SU(2)_{N+M}}$ is minimal with respect to the
Virasoro algebra extended by a conserved current of extra-symmetry\cite%
{ravanini_infinite_1988,Bagger_Virasoro_1988}. $N=1$ gives rise to the
series of unitary minimal models, while $N=2$ yields the series of the
unitary minimal superconformal models, where the Virasoro algebra is
extended to the super-Virasoro algebra by a conserved current with spin-$3/2$%
. This series of models has a superconformal symmetry, including the
critical point $D$ ($M=1$). For $N=3$, the Virasoro algebra is further
extended by a non-local conserved current with spin-$7/5$, and this series
of models has a fractional superconformal symmetry\cite%
{bernard_fractional_1990,argyres_structure_1991,chung_fractional_1992}, and
the tricritical point $C$ belongs to this family with $M=3$.

Notice that the fractional superconformal symmetry is only valid for the
two-dimensional wave function. The quantum self-duality is necessary for
the fractional superconformal symmetry. Via analyzing the topological anyon
sectors of the transfer operators inserted with matrix product operators\cite%
{Shadowofanyons,AnyonMPO}, the continuous topological phase transition along
the self-dual line is still driven by the condensation of $b$ anyons\cite%
{Bais2009} (see Supplementary Material). To the best of our knowledge, it is
for the first time to find such a tricritical point, and the quantum
criticality of this topological phase transition represents a new
universality class.

\textit{Summary and outlook}. - We have fully studied the non-abelian
topological phase transitions out of a generic DFib topological state. The key
point is the quantum self-duality. Along the self-dual line, three exactly
solvable points guide us to map out the global phase diagram. The tensor
network representation of the deformed DFib wave function on the square lattice
helps us determine the phase boundaries away from the self-dual line. More
importantly, a quantum tricritical point represents a new universality
class of quantum criticality, where the equal-time correlators are
characterized by a coset CFT and the wave function displays an emergent
fractional superconformal symmetry. Our method can be extended to study
topological phase transitions of other non-abelian topological phases.

A natural question is what relationship between the conformal quantum
criticality and the corresponding non-abelian topological phases. Another
issue is how the conformal quantum criticality will be changed when
the dynamics of the model Hamiltonian for the DFib topological phase is
considered. These problems are deserved further considerations. To realize the DFib
topological phase experimentally, it is known that parafermions appear in
the fractional quantum Hall effect at filling factors $\nu =1/m$ when two
counter-propagating edge states from two adjacent superconducting pairing
and spin-orbit induced tunneling\cite{NatureComm}, and then a two-dimensional
array of parafermions can support the Fibonacci anyons\cite{Alicea}.

\textit{Acknowledgment}.- The authors would like to thank Guo-Yi Zhu for his
stimulating discussions and acknowledge the support by the National Key
Research and Development Program of MOST of China (2016YFYA0300300).

\bibliography{ref}

\begin{thebibliography}{35}
\expandafter\ifx\csname natexlab\endcsname\relax\def\natexlab#1{#1}\fi
\expandafter\ifx\csname bibnamefont\endcsname\relax
  \def\bibnamefont#1{#1}\fi
\expandafter\ifx\csname bibfnamefont\endcsname\relax
  \def\bibfnamefont#1{#1}\fi
\expandafter\ifx\csname citenamefont\endcsname\relax
  \def\citenamefont#1{#1}\fi
\expandafter\ifx\csname url\endcsname\relax
  \def\url#1{\texttt{#1}}\fi
\expandafter\ifx\csname urlprefix\endcsname\relax\def\urlprefix{URL }\fi
\providecommand{\bibinfo}[2]{#2}
\providecommand{\eprint}[2][]{\url{#2}}

\bibitem[{\citenamefont{Kitaev}(2003)}]{kitaevToricCode2003}
\bibinfo{author}{\bibfnamefont{A.~Y.} \bibnamefont{Kitaev}},
  \bibinfo{journal}{Annals of Physics} \textbf{\bibinfo{volume}{303}},
  \bibinfo{pages}{2} (\bibinfo{year}{2003}), ISSN \bibinfo{issn}{0003-4916},
  \urlprefix\url{http://www.sciencedirect.com/science/article/pii/S0003491602000180}.

\bibitem[{\citenamefont{Freedman}(2003)}]{freedman2003}
\bibinfo{author}{\bibfnamefont{M.~H.} \bibnamefont{Freedman}},
  \bibinfo{journal}{Communications in Mathematical Physics}
  \textbf{\bibinfo{volume}{234}}, \bibinfo{pages}{129} (\bibinfo{year}{2003}).

\bibitem[{\citenamefont{Nayak et~al.}(2008)\citenamefont{Nayak, Simon, Stern,
  Freedman, and Das~Sarma}}]{RMP_TQC}
\bibinfo{author}{\bibfnamefont{C.}~\bibnamefont{Nayak}},
  \bibinfo{author}{\bibfnamefont{S.~H.} \bibnamefont{Simon}},
  \bibinfo{author}{\bibfnamefont{A.}~\bibnamefont{Stern}},
  \bibinfo{author}{\bibfnamefont{M.}~\bibnamefont{Freedman}}, \bibnamefont{and}
  \bibinfo{author}{\bibfnamefont{S.}~\bibnamefont{Das~Sarma}},
  \bibinfo{journal}{Rev. Mod. Phys.} \textbf{\bibinfo{volume}{80}},
  \bibinfo{pages}{1083} (\bibinfo{year}{2008}),
  \urlprefix\url{https://link.aps.org/doi/10.1103/RevModPhys.80.1083}.

\bibitem[{\citenamefont{Levin and Wen}(2005)}]{levin_string-net_2005}
\bibinfo{author}{\bibfnamefont{M.~A.} \bibnamefont{Levin}} \bibnamefont{and}
  \bibinfo{author}{\bibfnamefont{X.-G.} \bibnamefont{Wen}},
  \bibinfo{journal}{Physical Review B} \textbf{\bibinfo{volume}{71}}
  (\bibinfo{year}{2005}), ISSN \bibinfo{issn}{1098-0121, 1550-235X},
  \urlprefix\url{https://link.aps.org/doi/10.1103/PhysRevB.71.045110}.

\bibitem[{\citenamefont{Schulz et~al.}(2013)\citenamefont{Schulz, Dusuel,
  Schmidt, and Vidal}}]{J_Vidal_Golden_string_net}
\bibinfo{author}{\bibfnamefont{M.~D.} \bibnamefont{Schulz}},
  \bibinfo{author}{\bibfnamefont{S.}~\bibnamefont{Dusuel}},
  \bibinfo{author}{\bibfnamefont{K.~P.} \bibnamefont{Schmidt}},
  \bibnamefont{and} \bibinfo{author}{\bibfnamefont{J.}~\bibnamefont{Vidal}},
  \bibinfo{journal}{Phys. Rev. Lett.} \textbf{\bibinfo{volume}{110}},
  \bibinfo{pages}{147203} (\bibinfo{year}{2013}),
  \urlprefix\url{https://link.aps.org/doi/10.1103/PhysRevLett.110.147203}.

\bibitem[{\citenamefont{Dusuel and Vidal}(2015)}]{Vidal2015}
\bibinfo{author}{\bibfnamefont{S.}~\bibnamefont{Dusuel}} \bibnamefont{and}
  \bibinfo{author}{\bibfnamefont{J.}~\bibnamefont{Vidal}},
  \bibinfo{journal}{Phys. Rev. B} \textbf{\bibinfo{volume}{92}},
  \bibinfo{pages}{125150} (\bibinfo{year}{2015}),
  \urlprefix\url{https://link.aps.org/doi/10.1103/PhysRevB.92.125150}.

\bibitem[{\citenamefont{Mari\"en et~al.}(2017)\citenamefont{Mari\"en, Haegeman,
  Fendley, and Verstraete}}]{Condensation_driven}
\bibinfo{author}{\bibfnamefont{M.}~\bibnamefont{Mari\"en}},
  \bibinfo{author}{\bibfnamefont{J.}~\bibnamefont{Haegeman}},
  \bibinfo{author}{\bibfnamefont{P.}~\bibnamefont{Fendley}}, \bibnamefont{and}
  \bibinfo{author}{\bibfnamefont{F.}~\bibnamefont{Verstraete}},
  \bibinfo{journal}{Phys. Rev. B} \textbf{\bibinfo{volume}{96}},
  \bibinfo{pages}{155127} (\bibinfo{year}{2017}),
  \urlprefix\url{https://link.aps.org/doi/10.1103/PhysRevB.96.155127}.

\bibitem[{\citenamefont{Feiguin et~al.}(2007)\citenamefont{Feiguin, Trebst,
  Ludwig, Troyer, Kitaev, Wang, and Freedman}}]{GoldenChain}
\bibinfo{author}{\bibfnamefont{A.}~\bibnamefont{Feiguin}},
  \bibinfo{author}{\bibfnamefont{S.}~\bibnamefont{Trebst}},
  \bibinfo{author}{\bibfnamefont{A.~W.~W.} \bibnamefont{Ludwig}},
  \bibinfo{author}{\bibfnamefont{M.}~\bibnamefont{Troyer}},
  \bibinfo{author}{\bibfnamefont{A.}~\bibnamefont{Kitaev}},
  \bibinfo{author}{\bibfnamefont{Z.}~\bibnamefont{Wang}}, \bibnamefont{and}
  \bibinfo{author}{\bibfnamefont{M.~H.} \bibnamefont{Freedman}},
  \bibinfo{journal}{Phys. Rev. Lett.} \textbf{\bibinfo{volume}{98}},
  \bibinfo{pages}{160409} (\bibinfo{year}{2007}),
  \urlprefix\url{https://link.aps.org/doi/10.1103/PhysRevLett.98.160409}.

\bibitem[{\citenamefont{Trebst et~al.}(2008)\citenamefont{Trebst, Ardonne,
  Feiguin, Huse, Ludwig, and Troyer}}]{Trebst}
\bibinfo{author}{\bibfnamefont{S.}~\bibnamefont{Trebst}},
  \bibinfo{author}{\bibfnamefont{E.}~\bibnamefont{Ardonne}},
  \bibinfo{author}{\bibfnamefont{A.}~\bibnamefont{Feiguin}},
  \bibinfo{author}{\bibfnamefont{D.~A.} \bibnamefont{Huse}},
  \bibinfo{author}{\bibfnamefont{A.~W.~W.} \bibnamefont{Ludwig}},
  \bibnamefont{and} \bibinfo{author}{\bibfnamefont{M.}~\bibnamefont{Troyer}},
  \bibinfo{journal}{Phys. Rev. Lett.} \textbf{\bibinfo{volume}{101}},
  \bibinfo{pages}{050401} (\bibinfo{year}{2008}),
  \urlprefix\url{https://link.aps.org/doi/10.1103/PhysRevLett.101.050401}.

\bibitem[{\citenamefont{Gils et~al.}(2009)\citenamefont{Gils, Trebst, Kitaev,
  Ludwig, Troyer, and Wang}}]{gils_topology-driven_2009}
\bibinfo{author}{\bibfnamefont{C.}~\bibnamefont{Gils}},
  \bibinfo{author}{\bibfnamefont{S.}~\bibnamefont{Trebst}},
  \bibinfo{author}{\bibfnamefont{A.}~\bibnamefont{Kitaev}},
  \bibinfo{author}{\bibfnamefont{A.~W.~W.} \bibnamefont{Ludwig}},
  \bibinfo{author}{\bibfnamefont{M.}~\bibnamefont{Troyer}}, \bibnamefont{and}
  \bibinfo{author}{\bibfnamefont{Z.}~\bibnamefont{Wang}},
  \bibinfo{journal}{Nature Physics} \textbf{\bibinfo{volume}{5}},
  \bibinfo{pages}{834} (\bibinfo{year}{2009}), ISSN \bibinfo{issn}{1745-2473,
  1745-2481}, \urlprefix\url{http://www.nature.com/articles/nphys1396}.

\bibitem[{\citenamefont{Fendley}(2008)}]{fendley_topological_2008}
\bibinfo{author}{\bibfnamefont{P.}~\bibnamefont{Fendley}},
  \bibinfo{journal}{Annals of Physics} \textbf{\bibinfo{volume}{323}},
  \bibinfo{pages}{3113} (\bibinfo{year}{2008}), ISSN \bibinfo{issn}{00034916},
  \urlprefix\url{http://linkinghub.elsevier.com/retrieve/pii/S0003491608000614}.

\bibitem[{\citenamefont{Zhu and Zhang}(2019)}]{zhu_gapless_2019}
\bibinfo{author}{\bibfnamefont{G.-Y.} \bibnamefont{Zhu}} \bibnamefont{and}
  \bibinfo{author}{\bibfnamefont{G.-M.} \bibnamefont{Zhang}},
  \bibinfo{journal}{Phys. Rev. Lett.} \textbf{\bibinfo{volume}{122}},
  \bibinfo{pages}{176401} (\bibinfo{year}{2019}),
  \urlprefix\url{https://link.aps.org/doi/10.1103/PhysRevLett.122.176401}.

\bibitem[{\citenamefont{Fendley et~al.}(2013)\citenamefont{Fendley, Isakov, and
  Troyer}}]{fendley_fibonacci_2013}
\bibinfo{author}{\bibfnamefont{P.}~\bibnamefont{Fendley}},
  \bibinfo{author}{\bibfnamefont{S.~V.} \bibnamefont{Isakov}},
  \bibnamefont{and} \bibinfo{author}{\bibfnamefont{M.}~\bibnamefont{Troyer}},
  \bibinfo{journal}{Physical Review Letters} \textbf{\bibinfo{volume}{110}}
  (\bibinfo{year}{2013}), ISSN \bibinfo{issn}{0031-9007, 1079-7114},
  \urlprefix\url{https://link.aps.org/doi/10.1103/PhysRevLett.110.260408}.

\bibitem[{\citenamefont{Schuch et~al.}(2013)\citenamefont{Schuch, Poilblanc,
  Cirac, and P\'erez-Garc\'{\i}a}}]{PEPSTO}
\bibinfo{author}{\bibfnamefont{N.}~\bibnamefont{Schuch}},
  \bibinfo{author}{\bibfnamefont{D.}~\bibnamefont{Poilblanc}},
  \bibinfo{author}{\bibfnamefont{J.~I.} \bibnamefont{Cirac}}, \bibnamefont{and}
  \bibinfo{author}{\bibfnamefont{D.}~\bibnamefont{P\'erez-Garc\'{\i}a}},
  \bibinfo{journal}{Phys. Rev. Lett.} \textbf{\bibinfo{volume}{111}},
  \bibinfo{pages}{090501} (\bibinfo{year}{2013}),
  \urlprefix\url{https://link.aps.org/doi/10.1103/PhysRevLett.111.090501}.

\bibitem[{\citenamefont{Haegeman
  et~al.}(2015{\natexlab{a}})\citenamefont{Haegeman, Van~Acoleyen, Schuch,
  Cirac, and Verstraete}}]{GaugingTNS}
\bibinfo{author}{\bibfnamefont{J.}~\bibnamefont{Haegeman}},
  \bibinfo{author}{\bibfnamefont{K.}~\bibnamefont{Van~Acoleyen}},
  \bibinfo{author}{\bibfnamefont{N.}~\bibnamefont{Schuch}},
  \bibinfo{author}{\bibfnamefont{J.~I.} \bibnamefont{Cirac}}, \bibnamefont{and}
  \bibinfo{author}{\bibfnamefont{F.}~\bibnamefont{Verstraete}},
  \bibinfo{journal}{Phys. Rev. X} \textbf{\bibinfo{volume}{5}},
  \bibinfo{pages}{011024} (\bibinfo{year}{2015}{\natexlab{a}}),
  \urlprefix\url{https://link.aps.org/doi/10.1103/PhysRevX.5.011024}.

\bibitem[{\citenamefont{Wu}(1982)}]{wu_potts_1982}
\bibinfo{author}{\bibfnamefont{F.~Y.} \bibnamefont{Wu}},
  \bibinfo{journal}{Reviews of Modern Physics} \textbf{\bibinfo{volume}{54}},
  \bibinfo{pages}{235} (\bibinfo{year}{1982}), ISSN \bibinfo{issn}{0034-6861},
  \urlprefix\url{https://link.aps.org/doi/10.1103/RevModPhys.54.235}.

\bibitem[{\citenamefont{Fendley and Jacobsen}(2008)}]{fendley_critical_2008}
\bibinfo{author}{\bibfnamefont{P.}~\bibnamefont{Fendley}} \bibnamefont{and}
  \bibinfo{author}{\bibfnamefont{J.~L.} \bibnamefont{Jacobsen}},
  \bibinfo{journal}{Journal of Physics A: Mathematical and Theoretical}
  \textbf{\bibinfo{volume}{41}}, \bibinfo{pages}{215001}
  (\bibinfo{year}{2008}), ISSN \bibinfo{issn}{1751-8113, 1751-8121},
  \urlprefix\url{http://stacks.iop.org/1751-8121/41/i=21/a=215001?key=crossref.b622c38a08a1779c8dd3baf71daf5bba}.

\bibitem[{\citenamefont{Dotsenko et~al.}()\citenamefont{Dotsenko, Jacobsen,
  Lewis, and Picco}}]{dotsenko_coupled_1999}
\bibinfo{author}{\bibfnamefont{V.}~\bibnamefont{Dotsenko}},
  \bibinfo{author}{\bibfnamefont{J.~L.} \bibnamefont{Jacobsen}},
  \bibinfo{author}{\bibfnamefont{M.-A.} \bibnamefont{Lewis}}, \bibnamefont{and}
  \bibinfo{author}{\bibfnamefont{M.}~\bibnamefont{Picco}},
  \bibinfo{journal}{Nuclear Physics B} \textbf{\bibinfo{volume}{546}} (????),
  ISSN \bibinfo{issn}{0550-3213},
  \urlprefix\url{http://www.sciencedirect.com/science/article/pii/S0550321399000978}.

\bibitem[{\citenamefont{Gu et~al.}(2009)\citenamefont{Gu, Levin, Swingle, and
  Wen}}]{GuTensorNetwork}
\bibinfo{author}{\bibfnamefont{Z.-C.} \bibnamefont{Gu}},
  \bibinfo{author}{\bibfnamefont{M.}~\bibnamefont{Levin}},
  \bibinfo{author}{\bibfnamefont{B.}~\bibnamefont{Swingle}}, \bibnamefont{and}
  \bibinfo{author}{\bibfnamefont{X.-G.} \bibnamefont{Wen}},
  \bibinfo{journal}{Physical Review B} \textbf{\bibinfo{volume}{79}}
  (\bibinfo{year}{2009}), ISSN \bibinfo{issn}{1098-0121, 1550-235X},
  \urlprefix\url{https://link.aps.org/doi/10.1103/PhysRevB.79.085118}.

\bibitem[{\citenamefont{Buerschaper et~al.}(2009)\citenamefont{Buerschaper,
  Aguado, and Vidal}}]{buerschaper_explicit_2009}
\bibinfo{author}{\bibfnamefont{O.}~\bibnamefont{Buerschaper}},
  \bibinfo{author}{\bibfnamefont{M.}~\bibnamefont{Aguado}}, \bibnamefont{and}
  \bibinfo{author}{\bibfnamefont{G.}~\bibnamefont{Vidal}},
  \bibinfo{journal}{Physical Review B} \textbf{\bibinfo{volume}{79}}
  (\bibinfo{year}{2009}), ISSN \bibinfo{issn}{1098-0121, 1550-235X},
  \urlprefix\url{https://link.aps.org/doi/10.1103/PhysRevB.79.085119}.

\bibitem[{\citenamefont{Zanardi et~al.}(2007)\citenamefont{Zanardi, Giorda, and
  Cozzini}}]{fedility_2007}
\bibinfo{author}{\bibfnamefont{P.}~\bibnamefont{Zanardi}},
  \bibinfo{author}{\bibfnamefont{P.}~\bibnamefont{Giorda}}, \bibnamefont{and}
  \bibinfo{author}{\bibfnamefont{M.}~\bibnamefont{Cozzini}},
  \bibinfo{journal}{Phys. Rev. Lett.} \textbf{\bibinfo{volume}{99}},
  \bibinfo{pages}{100603} (\bibinfo{year}{2007}),
  \urlprefix\url{https://link.aps.org/doi/10.1103/PhysRevLett.99.100603}.

\bibitem[{\citenamefont{Fidkowski et~al.}(2009)\citenamefont{Fidkowski,
  Freedman, Nayak, Walker, and Wang}}]{fidkowski_string_2009}
\bibinfo{author}{\bibfnamefont{L.}~\bibnamefont{Fidkowski}},
  \bibinfo{author}{\bibfnamefont{M.}~\bibnamefont{Freedman}},
  \bibinfo{author}{\bibfnamefont{C.}~\bibnamefont{Nayak}},
  \bibinfo{author}{\bibfnamefont{K.}~\bibnamefont{Walker}}, \bibnamefont{and}
  \bibinfo{author}{\bibfnamefont{Z.}~\bibnamefont{Wang}},
  \bibinfo{journal}{Commun. Math. Phys.} \textbf{\bibinfo{volume}{287}},
  \bibinfo{pages}{805} (\bibinfo{year}{2009}), ISSN \bibinfo{issn}{1432-0916},
  \urlprefix\url{https://doi.org/10.1007/s00220-009-0757-9}.

\bibitem[{\citenamefont{Fendley and Fradkin}(2005)}]{Fendley_2005}
\bibinfo{author}{\bibfnamefont{P.}~\bibnamefont{Fendley}} \bibnamefont{and}
  \bibinfo{author}{\bibfnamefont{E.}~\bibnamefont{Fradkin}},
  \bibinfo{journal}{Phys. Rev. B} \textbf{\bibinfo{volume}{72}},
  \bibinfo{pages}{024412} (\bibinfo{year}{2005}),
  \urlprefix\url{https://link.aps.org/doi/10.1103/PhysRevB.72.024412}.

\bibitem[{\citenamefont{Ardonne et~al.}(2004)\citenamefont{Ardonne, Fendley,
  and Fradkin}}]{ardonne}
\bibinfo{author}{\bibfnamefont{E.}~\bibnamefont{Ardonne}},
  \bibinfo{author}{\bibfnamefont{P.}~\bibnamefont{Fendley}}, \bibnamefont{and}
  \bibinfo{author}{\bibfnamefont{E.}~\bibnamefont{Fradkin}},
  \bibinfo{journal}{Annals of Physics} \textbf{\bibinfo{volume}{310}},
  \bibinfo{pages}{493} (\bibinfo{year}{2004}), ISSN \bibinfo{issn}{00034916},
  \urlprefix\url{http://linkinghub.elsevier.com/retrieve/pii/S0003491604000247}.

\bibitem[{\citenamefont{Goodard et~al.}(1985)\citenamefont{Goodard, Kent, and
  Olive}}]{goodard_virasoro_1985}
\bibinfo{author}{\bibfnamefont{P.}~\bibnamefont{Goodard}},
  \bibinfo{author}{\bibfnamefont{A.}~\bibnamefont{Kent}}, \bibnamefont{and}
  \bibinfo{author}{\bibfnamefont{D.}~\bibnamefont{Olive}},
  \bibinfo{journal}{Physics Letters B} \textbf{\bibinfo{volume}{152}},
  \bibinfo{pages}{88} (\bibinfo{year}{1985}), ISSN \bibinfo{issn}{03702693},
  \urlprefix\url{http://linkinghub.elsevier.com/retrieve/pii/0370269385911451}.

\bibitem[{\citenamefont{Ravanini}(1988)}]{ravanini_infinite_1988}
\bibinfo{author}{\bibfnamefont{F.}~\bibnamefont{Ravanini}},
  \bibinfo{journal}{Modern Physics Letters A} \textbf{\bibinfo{volume}{03}},
  \bibinfo{pages}{397} (\bibinfo{year}{1988}), ISSN \bibinfo{issn}{0217-7323,
  1793-6632},
  \urlprefix\url{http://www.worldscientific.com/doi/abs/10.1142/S0217732388000490}.

\bibitem[{\citenamefont{Bagger et~al.}(1988)\citenamefont{Bagger, Nemeschansky,
  and Yankielowicz}}]{Bagger_Virasoro_1988}
\bibinfo{author}{\bibfnamefont{J.}~\bibnamefont{Bagger}},
  \bibinfo{author}{\bibfnamefont{D.}~\bibnamefont{Nemeschansky}},
  \bibnamefont{and}
  \bibinfo{author}{\bibfnamefont{S.}~\bibnamefont{Yankielowicz}},
  \bibinfo{journal}{Phys. Rev. Lett.} \textbf{\bibinfo{volume}{60}},
  \bibinfo{pages}{389} (\bibinfo{year}{1988}),
  \urlprefix\url{https://link.aps.org/doi/10.1103/PhysRevLett.60.389}.

\bibitem[{\citenamefont{Bernard and Leclair}(1990)}]{bernard_fractional_1990}
\bibinfo{author}{\bibfnamefont{D.}~\bibnamefont{Bernard}} \bibnamefont{and}
  \bibinfo{author}{\bibfnamefont{A.}~\bibnamefont{Leclair}},
  \bibinfo{journal}{Physics Letters B} \textbf{\bibinfo{volume}{247}},
  \bibinfo{pages}{309} (\bibinfo{year}{1990}), ISSN \bibinfo{issn}{0370-2693},
  \urlprefix\url{http://www.sciencedirect.com/science/article/pii/037026939090901H}.

\bibitem[{\citenamefont{Argyres et~al.}(1991)\citenamefont{Argyres,
  Grochocinski, and Henry~Tye}}]{argyres_structure_1991}
\bibinfo{author}{\bibfnamefont{P.~C.} \bibnamefont{Argyres}},
  \bibinfo{author}{\bibfnamefont{J.~M.} \bibnamefont{Grochocinski}},
  \bibnamefont{and} \bibinfo{author}{\bibfnamefont{S.~H.}
  \bibnamefont{Henry~Tye}}, \bibinfo{journal}{Nuclear Physics B}
  \textbf{\bibinfo{volume}{367}}, \bibinfo{pages}{217} (\bibinfo{year}{1991}),
  ISSN \bibinfo{issn}{0550-3213},
  \urlprefix\url{http://www.sciencedirect.com/science/article/pii/0550321391900483}.

\bibitem[{\citenamefont{Chung et~al.}(1992)\citenamefont{Chung, Lyman, and
  Tye}}]{chung_fractional_1992}
\bibinfo{author}{\bibfnamefont{S.-W.} \bibnamefont{Chung}},
  \bibinfo{author}{\bibfnamefont{E.}~\bibnamefont{Lyman}}, \bibnamefont{and}
  \bibinfo{author}{\bibfnamefont{S.-H.~H.} \bibnamefont{Tye}},
  \bibinfo{journal}{Int. J. Mod. Phys. A} \textbf{\bibinfo{volume}{07}},
  \bibinfo{pages}{3339} (\bibinfo{year}{1992}), ISSN \bibinfo{issn}{0217-751X},
  \urlprefix\url{https://www.worldscientific.com/doi/abs/10.1142/S0217751X92001484}.

\bibitem[{\citenamefont{Haegeman
  et~al.}(2015{\natexlab{b}})\citenamefont{Haegeman, Zauner, Schuch, and
  Verstraete}}]{Shadowofanyons}
\bibinfo{author}{\bibfnamefont{J.}~\bibnamefont{Haegeman}},
  \bibinfo{author}{\bibfnamefont{V.}~\bibnamefont{Zauner}},
  \bibinfo{author}{\bibfnamefont{N.}~\bibnamefont{Schuch}}, \bibnamefont{and}
  \bibinfo{author}{\bibfnamefont{F.}~\bibnamefont{Verstraete}},
  \bibinfo{journal}{NATURE COMMUNICATIONS} \textbf{\bibinfo{volume}{6}},
  \bibinfo{pages}{8} (\bibinfo{year}{2015}{\natexlab{b}}), ISSN
  \bibinfo{issn}{2041-1723},
  \urlprefix\url{http://dx.doi.org/10.1038/ncomms9284}.

\bibitem[{\citenamefont{Bultinck et~al.}(2017)\citenamefont{Bultinck,
  Mari{\"e}n, Williamson, {\c S}ahino{\u g}lu, Haegeman, and
  Verstraete}}]{AnyonMPO}
\bibinfo{author}{\bibfnamefont{N.}~\bibnamefont{Bultinck}},
  \bibinfo{author}{\bibfnamefont{M.}~\bibnamefont{Mari{\"e}n}},
  \bibinfo{author}{\bibfnamefont{D.~J.} \bibnamefont{Williamson}},
  \bibinfo{author}{\bibfnamefont{M.~B.} \bibnamefont{{\c S}ahino{\u g}lu}},
  \bibinfo{author}{\bibfnamefont{J.}~\bibnamefont{Haegeman}}, \bibnamefont{and}
  \bibinfo{author}{\bibfnamefont{F.}~\bibnamefont{Verstraete}},
  \bibinfo{journal}{Annals of Physics} \textbf{\bibinfo{volume}{378}},
  \bibinfo{pages}{183} (\bibinfo{year}{2017}), ISSN \bibinfo{issn}{0003-4916},
  \urlprefix\url{http://www.sciencedirect.com/science/article/pii/S0003491617300040}.

\bibitem[{\citenamefont{Bais and Slingerland}(2009)}]{Bais2009}
\bibinfo{author}{\bibfnamefont{F.~A.} \bibnamefont{Bais}} \bibnamefont{and}
  \bibinfo{author}{\bibfnamefont{J.~K.} \bibnamefont{Slingerland}},
  \bibinfo{journal}{Phys. Rev. B} \textbf{\bibinfo{volume}{79}},
  \bibinfo{pages}{045316} (\bibinfo{year}{2009}),
  \urlprefix\url{https://link.aps.org/doi/10.1103/PhysRevB.79.045316}.

\bibitem[{\citenamefont{Clarke et~al.}(2013)\citenamefont{Clarke, Alicea, and
  Shtengel}}]{NatureComm}
\bibinfo{author}{\bibfnamefont{D.~J.} \bibnamefont{Clarke}},
  \bibinfo{author}{\bibfnamefont{J.}~\bibnamefont{Alicea}}, \bibnamefont{and}
  \bibinfo{author}{\bibfnamefont{K.}~\bibnamefont{Shtengel}},
  \bibinfo{journal}{Nature Communications} \textbf{\bibinfo{volume}{4}},
  \bibinfo{pages}{1348} (\bibinfo{year}{2013}).

\bibitem[{\citenamefont{Mong et~al.}(2014)\citenamefont{Mong, Clarke, Alicea,
  Lindner, Fendley, Nayak, Oreg, Stern, Berg, Shtengel et~al.}}]{Alicea}
\bibinfo{author}{\bibfnamefont{R.~S.~K.} \bibnamefont{Mong}},
  \bibinfo{author}{\bibfnamefont{D.~J.} \bibnamefont{Clarke}},
  \bibinfo{author}{\bibfnamefont{J.}~\bibnamefont{Alicea}},
  \bibinfo{author}{\bibfnamefont{N.~H.} \bibnamefont{Lindner}},
  \bibinfo{author}{\bibfnamefont{P.}~\bibnamefont{Fendley}},
  \bibinfo{author}{\bibfnamefont{C.}~\bibnamefont{Nayak}},
  \bibinfo{author}{\bibfnamefont{Y.}~\bibnamefont{Oreg}},
  \bibinfo{author}{\bibfnamefont{A.}~\bibnamefont{Stern}},
  \bibinfo{author}{\bibfnamefont{E.}~\bibnamefont{Berg}},
  \bibinfo{author}{\bibfnamefont{K.}~\bibnamefont{Shtengel}},
  \bibnamefont{et~al.}, \bibinfo{journal}{Phys. Rev. X}
  \textbf{\bibinfo{volume}{4}}, \bibinfo{pages}{011036} (\bibinfo{year}{2014}),
  \urlprefix\url{https://link.aps.org/doi/10.1103/PhysRevX.4.011036}.

\end{thebibliography}

\begin{widetext}

\section{Supplemental Material}

\subsection{Deformed wave function and its norm in the loop basis}

In the main text, we have the orthogonal local quantum states $|1\rangle $
and $|\tau \rangle $ on the original lattice and the orthogonal local
quantum states $|\hat{1}\rangle $ and $|\hat{\tau}\rangle $ on the dual
lattice. They are related by the following transformation
\begin{equation}
\left[
\begin{array}{c}
|\hat{1}\rangle \\
|\hat{\tau}\rangle%
\end{array}%
\right] =F\left[
\begin{array}{c}
|1\rangle \\
|\tau \rangle%
\end{array}%
\right] =\frac{1}{\phi }\left[
\begin{array}{cc}
1 & \sqrt{\phi } \\
\sqrt{\phi } & 1%
\end{array}%
\right] \left[
\begin{array}{c}
|1\rangle \\
|\tau \rangle%
\end{array}%
\right] .  \label{transformation}
\end{equation}%
It has been proved that the the quantum net wave function takes the same
form when rewritten on the dual lattice in the representation of $|\hat{1}%
\rangle $ and $|\hat{\tau}\rangle $, exhibiting the quantum self-duality.
This fact can be understood from the completely packed quantum loop model,
from which the Fendley's quantum net originates. The basis for the quantum
loop model is given by the completely packed loop configuration $\mathcal{L}$
consisting of two non-orthogonal states $|\tau \rangle $ and $|\hat{\tau}%
\rangle $, see Fig. \ref{CPL3} (a). Their overlaps are determined as
\begin{equation}
\left[
\begin{array}{cc}
\langle \tau |\tau \rangle & \langle \tau |\hat{\tau}\rangle \\
\langle \hat{\tau}|\tau \rangle & \langle \hat{\tau}|\bar{\tau}\rangle%
\end{array}%
\right] =\left[
\begin{array}{cc}
1 & \lambda \\
\lambda & 1%
\end{array}%
\right] .  \label{overlap}
\end{equation}%
According to Eq.(\ref{transformation}), the value of $\lambda $ is
determined to $-1/\phi $. So the loop configurations $|\mathcal{L}\rangle $ are
non-orthogonal. In the following, we will show that the value of $\lambda $
can be changed by deforming the wave function. In the non-orthogonal loop
basis, the quantum net wave function can be expressed as:
\begin{equation}
|\Psi \rangle =\sum_{\mathcal{L}}\phi ^{n_{\mathcal{L}}}|\mathcal{L}\rangle ,
\label{quantum_loop}
\end{equation}%
where $n_{\mathcal{L}}$ is the number of loops in $\mathcal{L}$ and the
loop fugacity is $\phi $. Under the quantum duality where the $|\tau \rangle $
and $|\hat{\tau}\rangle $ are exchanged with each other, the loop lattice as
the medial lattice of the net lattice and dual net lattice is unchanged, so
it is obvious that the wave function (\ref{quantum_loop}) is invariant.
Therefore, when Eq. (\ref{quantum_loop}) is expressed in the original
lattice on the basis $|1\rangle $ and $|\tau \rangle $ and the dual square
lattice on the basis $|\hat{1}\rangle $ and $|\hat{\tau}\rangle $, the
wavefunction should have the same form.
\begin{figure}[tbp]
\centering
\includegraphics[width=12cm,trim=0 180 10 10,clip]{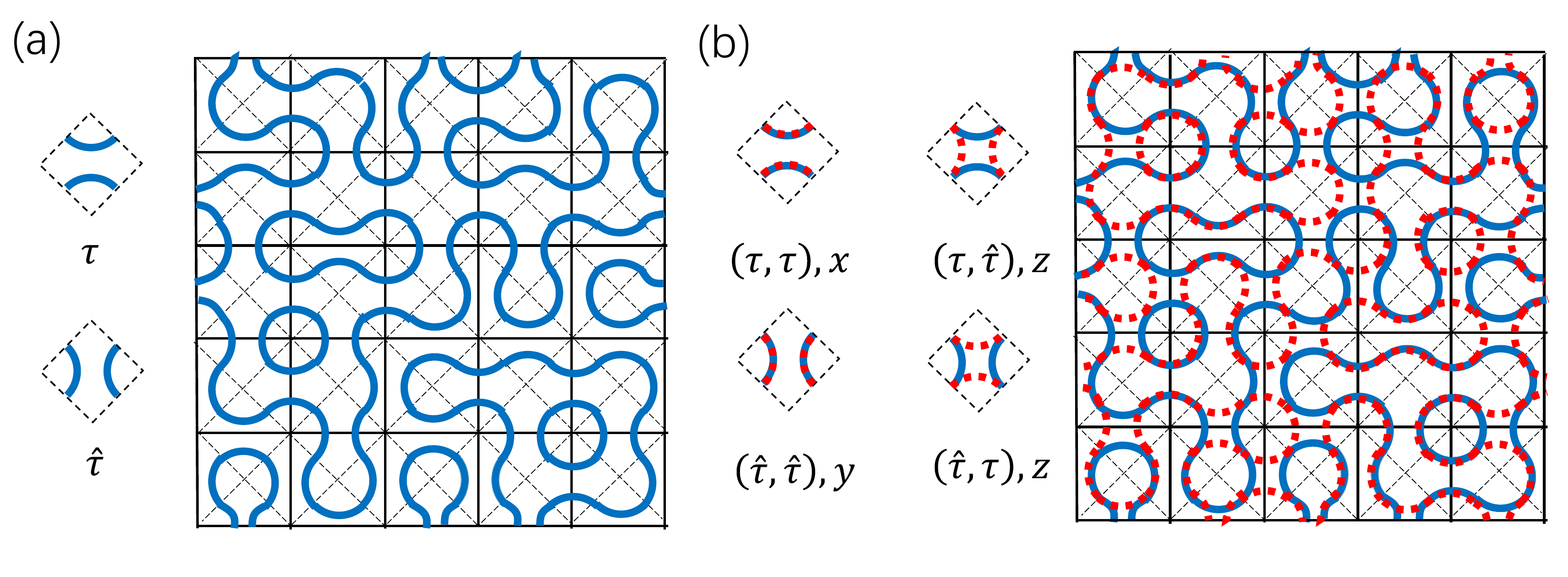}
\caption{(a) A typical completely packed loop basis $|\mathcal{L}\rangle $ consists
of the local non-orthogonal states $|\protect\tau \rangle $ and $|\hat{%
\protect\tau}\rangle $, the left and right, up and down sides are connected
periodically such that all loops in the configuration are closed. The black
full lines are net lattice, the black dash lines are loop lattice, and the
blue thick lines are completely packed loop configuration. (b) A typical
configuration $(\mathcal{L},\mathcal{L}^{\prime })$ of two-coupled
completely packed loop model consists of the four local configurations. $x$,
$y$ and $z$ are local Boltzmann weights for these local configurations. The
blue full lines stand for the configuration $\mathcal{L}$ form ket, the the
red dash lines stand for the configuration $\mathcal{L}^{\prime }$ form bra.
}
\label{CPL3}
\end{figure}

Actually the norm of the deformed wave function can be expressed in the
non-orthogonal loop basis:
\begin{equation}
\mathcal{Z}=\langle \Psi (h,\hat{h})|\Psi (h,\hat{h})\rangle =\sum_{\mathcal{%
L},\mathcal{L}^{\prime }}\phi ^{n_{\mathcal{L}}+n_{\mathcal{L}^{\prime
}}}x^{n_{\tau \tau }}y^{n_{\hat{\tau}\hat{\tau}}}z^{n_{\tau \hat{\tau}}+n_{%
\hat{\tau}\tau }},  \label{loop_norm}
\end{equation}%
where $\mathcal{L}$ and $\mathcal{L}^{\prime }$ are the loop configurations
coming from ket and bra separately and the loop fugacity is the global part
of Boltzmann weights. $x$, $y$ and $z$ are the local Boltzmann weights for
four local configurations in Fig. \ref{CPL3} (b), while $n_{i,j}$ with $%
i,j=\tau ,\hat{\tau}$ is the total number of local configurations $(i,j)$ in
the global configuration $\left( \mathcal{L},\mathcal{L}^{\prime }\right) $,
see Fig.\ref{CPL3} (b). The values of $x,y$ and $z$ are determined by the
matrix elements of the squared deformation matrix $P^{2}$ in non-orthogonal
basis:
\begin{eqnarray}
x &=&\langle \tau |P^{2}(h,\hat{h})|\tau \rangle =\left( 1-h+\hat{h}\right)
^{2}-\frac{4\hat{h}}{\phi ^{2}}\left( 1-h\right) ,  \notag
\label{Boltzmann_weight} \\
y &=&\langle \hat{\tau}|P^{2}(h,\hat{h})|\hat{\tau}\rangle =\left( 1+h-\hat{h%
}\right) ^{2}-\frac{4h}{\phi ^{2}}\left( 1-\hat{h}\right) ,,  \notag \\
z &=&\langle \tau |P^{2}(h,\hat{h})|\hat{\tau}\rangle =\langle \hat{\tau}%
|P^{2}(h,\hat{h})|\tau \rangle =-\frac{1}{\phi }\left( 1-h-\hat{h}\right)
^{2}+\frac{4h\hat{h}}{\phi ^{2}}.
\end{eqnarray}%
Then the wave function norm can be regarded as the partition function of a
two-coupled loop model. On the self-dual line $h=\hat{h}$, we have
\begin{equation}
x=y=1-\frac{4h}{\phi ^{2}}\left( 1-h\right) ,\text{ }z=-\frac{1}{\phi }%
\left( 1-2h\right) ^{2}+\frac{4h^{2}}{\phi ^{2}},
\end{equation}%
and the partition function (\ref{loop_norm}) is simplified as
\begin{equation}
\mathcal{Z}=\left\langle \Psi \left( h,h\right) |\Psi \left( h,h\right)
\right\rangle \propto \sum_{\mathcal{L},\mathcal{L}^{\prime }}\phi ^{n_{%
\mathcal{L}}+n_{\mathcal{L}^{\prime }}}\lambda ^{n_{\tau \hat{\tau}}+n_{\hat{%
\tau}\tau }}.  \label{norm_loop}
\end{equation}%
with
\begin{equation}
\lambda =\frac{z}{x}=\frac{4h^{2}-\left( 2h-1\right) ^{2}\phi }{\left(
2h-1\right) ^{2}+\phi },  \label{parameter_transformation}
\end{equation}%
which is precisely the overlap between two non-orthogonal local states in
Eq.(\ref{overlap}). This fact can be easily understood from the local
configurations $(\tau ,\hat{\tau})$ and $(\hat{\tau},\tau )$ in Fig. \ref%
{CPL3} (b). So we find that, the effect of tuning the filter strength along
the self-dual line $h=\hat{h}$ is equivalent to changing the overlap $%
\langle \tau |\hat{\tau}\rangle =\langle \hat{\tau}|\tau \rangle $. The
wavefunction norm on the self-dual line in Eq. (\ref{norm_loop}) is nothing
but the self-dual coupled loop model. The
range of parameter on the self-dual line is limited to $h\in \left[ 0,%
\frac{\phi }{2}\right] $, corresponding to the range of $\lambda \in \left[
-1/\phi ,1\right] $.

In the loop representation, the decoupled point $D$ on the self-dual line is
obviously when $\lambda =1$ or $h=\frac{\phi }{2}$, and the partition
function is $\mathcal{Z}\propto \sum_{\mathcal{L},\mathcal{L}^{\prime }}\phi
^{n_{\mathcal{L}}+n_{\mathcal{L}^{\prime }}}$. At this point each loop
model is critical. However, the strongly coupled point $S$ is given by $%
\lambda =0$ corresponding to $h=\frac{\phi ^{2}}{2}\left( 1-\sqrt{\phi }%
\right) $ and $\mathcal{Z}\propto \sum_{\mathcal{L}}\phi ^{2n_{\mathcal{L}%
}} $. This loop model is non-critical because the fugacity $\phi ^{2}>2$. In
addition to these special points, it is pointed out that there is a new
critical point $C$ on the self-dual line. The transfer operator of the
self-dual coupled Potts model can be written in terms of a special $%
SO(4)_{3}$ Birman-Murakami-Wenzl algebra. By the
level-rank duality of the algebra, the position of the new critical point is
determined by $\lambda =-\sqrt{2}\sin \frac{\pi }{20}$ corresponding to
$h\approx 0.196952$.

\subsection{Tensor network representation of the Fendley's quantum net}

So far the quantum net wave function on a square lattice have not been written
in the tensor network representation, because the chromatic polynomial as a
non-local topological part of the ground state weight is difficult to deal with.
Here we would like to construct a triple-line tensor network state (TNS) for
the quantum net on a square lattice.

First of all, we start from the known TNS for the string-net model on a honeycomb
lattice. The physical degrees of freedom located at the edges has two
orthogonal quantum states labelled by $|1\rangle $ and $|\tau \rangle $, as
shown in Fig.\ref{nets} (a). The branching symbol $\delta _{ijk}$ stemming
from the non-Abelian fusion rule imposes the branching rule on the vertices.
It allows the following five configurations shown in Fig.\ref{nets} (a),
corresponding to the non-zero elements of $\delta _{ijk}$:
\begin{equation}
\delta _{111}=\delta _{1\tau \tau }=\delta _{\tau 1\tau }=\delta _{\tau \tau
1}=\delta _{\tau \tau \tau }=1.
\end{equation}%
Therefore, in the ground state wavefunction, $|\tau \rangle $ states form a
closed trivalent net $\mathcal{N}$ consisting of $\tau $ strings, a typical
net is displayed in Fig.\ref{nets} (b).
\begin{figure}[tbp]
\centering
\includegraphics[width=9cm,trim=0 10 0 0,clip]{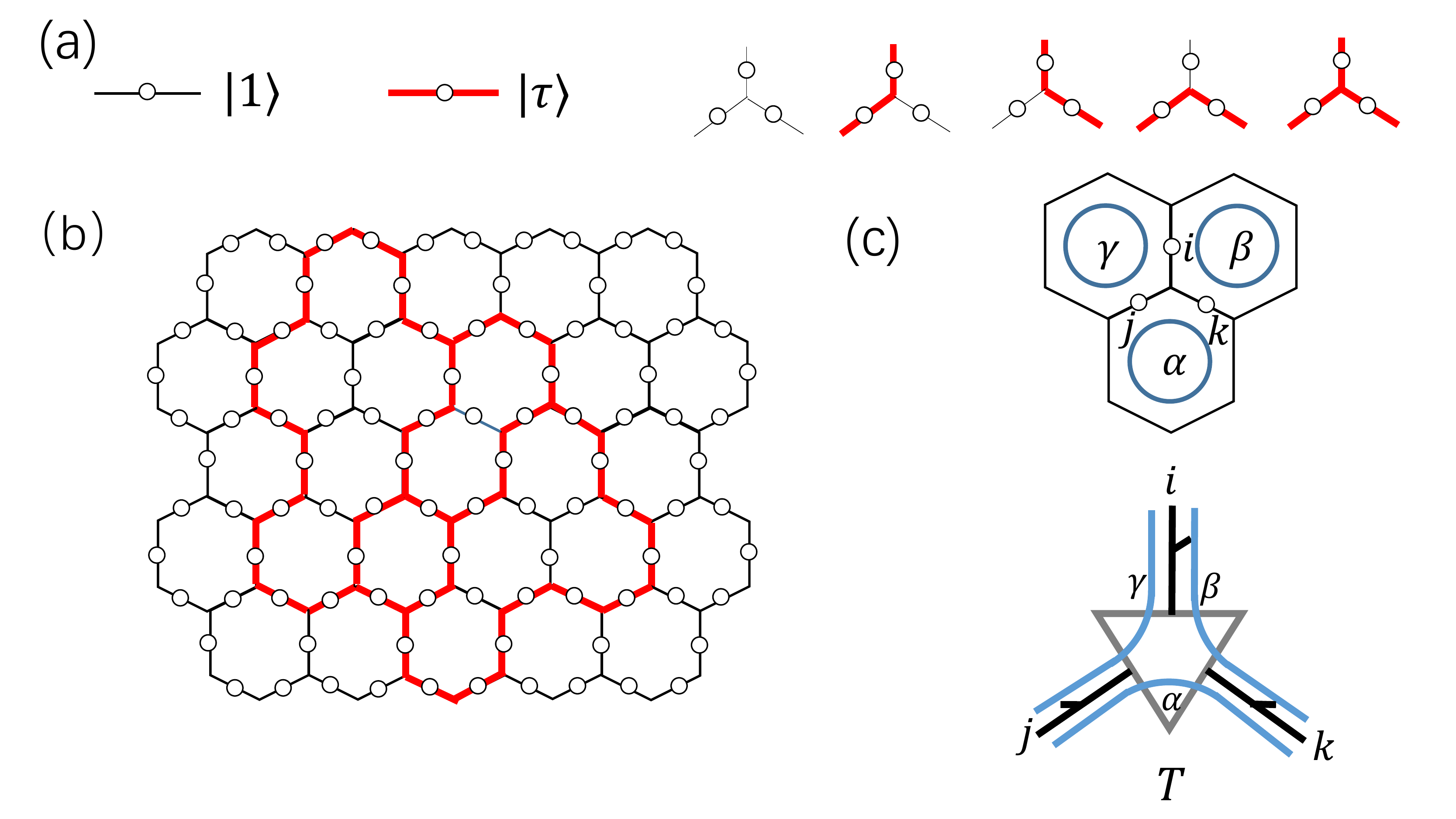}
\caption{(a) Two orthogonal quantum states $|1\rangle$ and $|\protect\tau%
\rangle$ on an edge, and the five local configurations allowed by the branching rule
on a vertex of trivalent lattice. The circles represent the physical degrees
of freedom. (b) A typical net configuration on a honeycomb lattice, the $%
\protect\tau$ strings formed by $|\protect\tau\rangle$ states are closed and
allowed to branch. (c) Up panel: The physical indices $i,j,k$ of the local tensor
locate at the edges and auxiliary indices $t,s,u$ live in the plaquette of the
lattice. Lower panel: The triplet-line local tensor $T^{ijk}_{tsu}$ on the
vertex of the honeycomb lattice for the string-net ground state. The legs of the
tensor connected by lines with only one index for simplicity are locked by
delta function, and the legs pointed out side of the paper are physical indices.}
\label{nets}
\end{figure}

With the chromatic polynomial, the ground state wavefunction for the
Fibonacci string-net can be explicitly expressed as
\begin{equation}
|\Psi _{\text{SN}}\rangle =\sum_{\mathcal{N}}\phi ^{3t_{\mathcal{N}}/4}\chi
_{\hat{\mathcal{N}}}\left( \phi ^{2}\right) |\mathcal{N}\rangle ,
\label{string_net}
\end{equation}%
where $t_{\mathcal{N}}$ is the number of vertices in the net $\mathcal{N}$ and
three $\tau $ strings meet together. So $\phi ^{3/4}$ can be viewed as
the vertex fugacity. The chromatic polynomial is non-local, and it only
relates to the topological properties of the net $\mathcal{N}$. To represent
the non-local wavefunction locally, the auxiliary degrees of freedom have to
be introduced, and the wavefunction can be decomposed into a local
structure, i.e., the triplet-line tensor network states (TNS):
\begin{equation}
|\Psi _{\text{SN}}\rangle =\sum_{\{ijk\}}\text{tTr}\left[ \underset{\text{%
vertex}}{\bigotimes }T_{\alpha \beta \gamma }^{ijk}\right] |\cdots ijk\cdots
\rangle ,  \label{string-net TNS}
\end{equation}%
where $T_{\alpha \beta \gamma }^{ijk}=\left( a_{\alpha }a_{\beta }a_{\gamma
}\right) ^{1/6}\sqrt{v_{i}v_{j}v_{k}}G_{\alpha \beta \gamma }^{ijk}$ is the
local tensor for the string-net model, and "tTr" denotes the contraction
over all virtual indices $\alpha $, $\beta $, $\gamma $. $G$, $v$ and $a$
tensors are uniquely determined by the Fibonacci topological order, and $%
v_{i}=\sqrt{d_{i}}$, $a_{i}=d_{i}/\left( \sum_{i}d_{i}^{2}\right) $ with $%
d_{1}=1$, $d_{\tau }=\phi $ as the quantum dimensions. Moreover, $G_{\alpha
\beta \gamma }^{ijk}=\delta _{ij\gamma }\delta _{\alpha \beta \gamma }\delta
_{i\beta k}\delta _{j\alpha k}\frac{F_{\alpha \beta \gamma }^{ijk}}{%
v_{k}v_{\gamma }}$, and the $F$-symbol is given by:
\begin{equation}
F=\left(
\begin{array}{cc}
F_{\tau \tau 1}^{\tau \tau 1} & F_{\tau \tau \tau }^{\tau \tau 1} \\
F_{\tau \tau 1}^{\tau \tau \tau } & F_{\tau \tau \tau }^{\tau \tau \tau }%
\end{array}%
\right) =\frac{1}{\phi }\left(
\begin{array}{cc}
1 & \sqrt{\phi } \\
\sqrt{\phi } & -1%
\end{array}%
\right) .  \label{fusion_matrix}
\end{equation}%
The graphical representation of the $T$ tensor is shown in Fig.\ref{nets}(c).

In the main text, the wavefunction of quantum net is defined by
\begin{equation}
|\Psi \rangle =\sum_{\mathcal{N}}\phi ^{\left. -L_{\mathcal{N}}\right/
2}\chi _{\hat{\mathcal{N}}}\left( \phi ^{2}\right) |\mathcal{N}\rangle ,
\label{Fendley_wavefunction}
\end{equation}%
where $L_{\mathcal{N}}$ is total length of $\tau $ string in net $\mathcal{N}
$. Note that the difference between string-net wavefunctions in Eq. (\ref%
{string_net}) and quantum net in Eq. (\ref{Fendley_wavefunction}) is the local
parts of the ground state weights and the types of the lattice. The local parts
(string tension and vertex fugacity) are very easy to deal with, so the key
point is to represent the chromatic polynomials for nets on the square
lattice in terms of tensor networks. From Eqs. (\ref{string_net}) and (\ref%
{string-net TNS}), we notice the correspondence between the chromatic
polynomials for nets on the honeycomb lattice and the tensor networks is
\begin{equation}
\chi _{\hat{\mathcal{N}}_{\text{hyc}}}\left( \phi ^{2}\right)
\Leftrightarrow \text{tTr}\left( \underset{\text{vertex}}{\bigotimes }\tilde{%
T}_{\alpha \beta \gamma }^{ijk}\right) ,  \label{correspondence}
\end{equation}%
where \textquotedblleft hyc\textquotedblright\ stands for the honeycomb
lattice and $\tilde{T}_{tsu}^{ijk}=\phi ^{-\frac{3}{4}\delta _{i,j}\delta
_{j,k}\delta _{k,\tau }}T_{tsu}^{ijk}$. Since the ground states are
degenerate, two above expressions might not be equal.

Thanks to the equations among the chromatic polynomials shown in Fig. \ref{derive_tensor} (a), we can
obtain the relation between $\chi _{\hat{\mathcal{N}}_{\text{hyc}}}(Q)$ and $%
\chi _{\hat{\mathcal{N}}_{\text{sq}}}(Q)$. More specifically, we can
transform the square lattice into a honeycomb lattice by splitting all the
sites and adding new edges, as shown in Fig. \ref{derive_tensor}. There
exist two different ways of splitting a single site, but the results are
equivalent to each other. A new created edge can be either occupied by $\tau
$ string or unoccupied. For a net on the square lattice, it might include
the bivalent, trivalent and tetra-valent vertices. After the lattice
transformation, there is no ambiguity for both bivalent and trivalent
vertices, because the chromatic polynomials for open nets are zero. For a
net with one tetra-valent vertex, its chromatic polynomial becomes a sum of
two chromatic polynomials on the honeycomb lattice, as shown in Fig. \ref%
{derive_tensor} (b). If a net on the square lattice has $n$ tetra-valent
vertices, its chromatic polynomial is a sum over $2n$ chromatic polynomials
on the honeycomb lattice.
\begin{figure}[tbp]
\centering
\includegraphics[width=8.5 cm,,trim=0 0 0 0,clip]{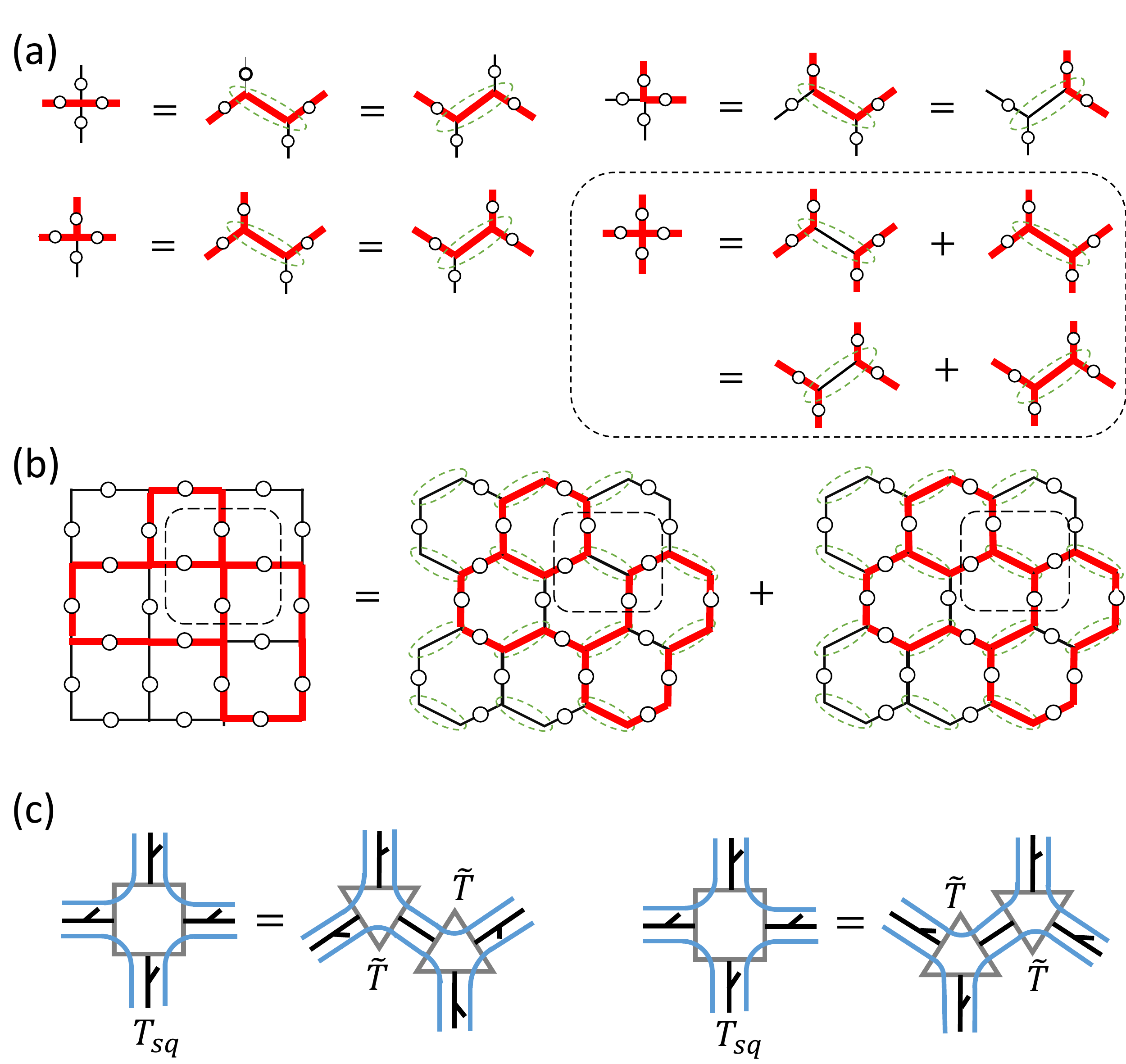}
\caption{(a) The square lattice can be transformed to a honeycomb lattice by
splitting a lattice site into two sites and adding new edges (in the green ovals),
notice that on the new edges there is no physical degree of freedom. There
are two equivalent ways of doing this. The single $\protect\tau $ string on
the vertex of square lattice is transformed into a single $\protect\tau $
string on the vertices of the honeycomb lattice, displayed by the first three
equations. The fourth equation in the box shows that the tetra-valent vertex
on a square lattice where two $\protect\tau $ stings meet can be reduced to
the sum of two graphs on the honeycomb lattice. (b) The chromatic polynomial for
a square lattice net with a tetra-valent vertex (in the box with dash line)
is equal to the sum of two chromatic polynomials for the honeycomb lattice nets.
(c) Contracting two tensors $\tilde{T}$ along one bond direction of the string-net TNS,
we obtain the tensor $T_{sq}$ for Fendley's wavefunction. Because there two equivalent
ways of transforming the vertex of a square lattice into two vertices of the
honeycomb lattice, there are two kinds of $T_{sq}$ tensors. Notice that the
physical index between two $\tilde{T}$ tensors belonging to the new edges is summed.}
\label{derive_tensor}
\end{figure}

Conversely, for the wave function with the weight $\chi _{\hat{\mathcal{N}}_{%
\text{hyc}}}(\phi ^{2})$ on the honeycomb lattice, summing over physical
degrees of freedom on the new edges (NE) of the honeycomb lattice
yields the wave function weighted by $\chi _{\hat{\mathcal{N}}_{\text{sq}%
}}(\phi ^{2})$ on the square lattice:
\begin{equation}
\sum_{\mathcal{N}_{\text{sq}}}\chi _{\hat{\mathcal{N}}_{\text{sq}}}\left(
\phi ^{2}\right) |\mathcal{N}_{\text{sq}}\rangle =\left[ \underset{\text{NE}}%
{\bigotimes }\sum_{i=1,\tau }\langle i|\right] \left[ \sum_{\mathcal{N}_{%
\text{hyc}}}\chi _{\hat{\mathcal{N}}_{\text{hyc}}}\left( \phi ^{2}\right) |%
\mathcal{N}_{\text{hyc}}\rangle \right] .  \label{relation}
\end{equation}%
This is the relation between $\chi _{\hat{\mathcal{N}}_{\text{hyc}}}(Q)$ and
$\chi _{\hat{\mathcal{N}}_{\text{sq}}}(Q)$, which only involves some local
operations. So the connection between the local tensors for a string-net and
local tensors for a quantum net is clear. Although the TNS representation
and chromatic polynomial representation of the ground state might not be
equal, the two expressions can not be distinguished locally. The TNSs for a
string-net on the honeycomb lattice and the corresponding quantum net on the
square lattice obey the same relation:
\begin{equation}
\sum_{\{ijkl\cdots \}}\text{tTr}\left( \underset{\text{vertex}}{\bigotimes
}T_{sq}\right) |\ ijkl\cdots \rangle
=\left[ \underset{\text{ne}}{\bigotimes }\sum_{i=1}^{\tau }\langle i|%
\right] \left[ \sum_{\{ijk\cdots \}}\text{tTr}\left( \underset{\text{vertex}}%
{\bigotimes }\tilde{T}\right) |\cdots ijk\cdots \rangle \right] .
\end{equation}
In this TNS, the sum over physical degrees of freedom on new edges is
equivalent to that there is no physical degrees on these new edges. Hence, the
local tensor $T_{\text{sq}}$ for a square lattice quantum net is given by
combining two tensors $\tilde{T}$ along one bond direction of the honeycomb
lattice, as displayed in Fig. \ref{derive_tensor} (c). Taking the string
tension into account, the TNS for the quantum net on the square lattice is
thus given by
\begin{equation}
|\Psi \rangle =\sum_{\{ijkl\cdots \}}\text{tTr}\left( \underset{\text{vertex}%
}{\bigotimes }\phi ^{-\frac{i+j+k+l}{4}}T_{\text{sq}}\right) |ijkl\cdots
\rangle ,
\end{equation}%
where the string tension on each tensor reduces to $\phi ^{1/4}$.

\subsection{Transfer operator and matrix product operator}

In order to detect the mechanism of the quantum phase transition, we
analysis the fate of anyons with matrix product operator (MPO). The MPO acts
on virtual indices of the tensor networks, satisfying the pulling through
condition, so the invisible MPO can be moved freely. Although our TNS for the quantum net is
different from the string-net TNS, their MPOs are the same. The reason is
that the local tensor $T_{\text{sq}}$ comes from local tensor $T$ of the
string-net, we only do some local operations on the physical degrees of
freedom of the tensor, while the auxiliary degrees of freedom that MPO acts
on remain unchanged. This is also an evidence that the quantum net on the
square lattice and the string-net have the same topological order, because
the MPO algebra is a necessary condition for topologically ordered TNS.

The local tensor of MPO is also determined by $G$ and $v$ tensors. There are
two local tensors
\begin{equation}
B_{1,(jl),(im)}^{(jki),(lkm)}=G_{lm1}^{ijk}\sqrt{v_{i}v_{j}v_{l}v_{m}}\text{%
, \ }B_{\tau ,(jl),(im)}^{(jki),(lkm)}=G_{lm\tau }^{ijk}\sqrt{%
v_{i}v_{j}v_{l}v_{m}},
\end{equation}%
where the superscribes are the physical indices and the subscribes are
virtual indices, as displayed in the right panel of Fig. \ref%
{Transfer_Operator} (a). The MPOs generated by $B_{1}$ and $B_{\tau }$ are
given by
\begin{eqnarray}
O_{1} &=&\sum_{\{(jki),(lkm)\}}\text{tr}\left( \prod_{l}B_{1}^{\left(
j_{l}k_{l}i_{l}\right) ,\left( l_{1}k_{1}m_{1}\right) }\right)
|\{jki\}\rangle \langle \{lkm\}|,  \notag \\
O_{\tau } &=&\sum_{\{(jki),(lkm)\}}\text{tr}\left( \prod_{l}B_{\tau
}^{\left( j_{l}k_{l}i_{l}\right) ,\left( l_{1}k_{1}m_{1}\right) }\right)
|\{jki\}\rangle \langle \{lkm\}|.
\end{eqnarray}%
$O_{1}$ is trivial in the sense that it is equivalent to an identity
operator, while $O_{\tau }$ is non-trivial, because it has two eigenvalues $%
\phi $ and $-\phi ^{-1}$. Both $O_{1}$ and $O_{\tau }$ satisfy the Fibonacci
fusion rule. Notice that these two MPOs are unchanged for different values
of $h$ and $\hat{h}$. To investigate the mechanism of the phase transition,
we need to systematically classify the topological sectors of the tensor
networks with MPOs.
\begin{figure}[tbp]
\centering
\includegraphics[width=8.5cm, trim=0 0 0 0,clip]{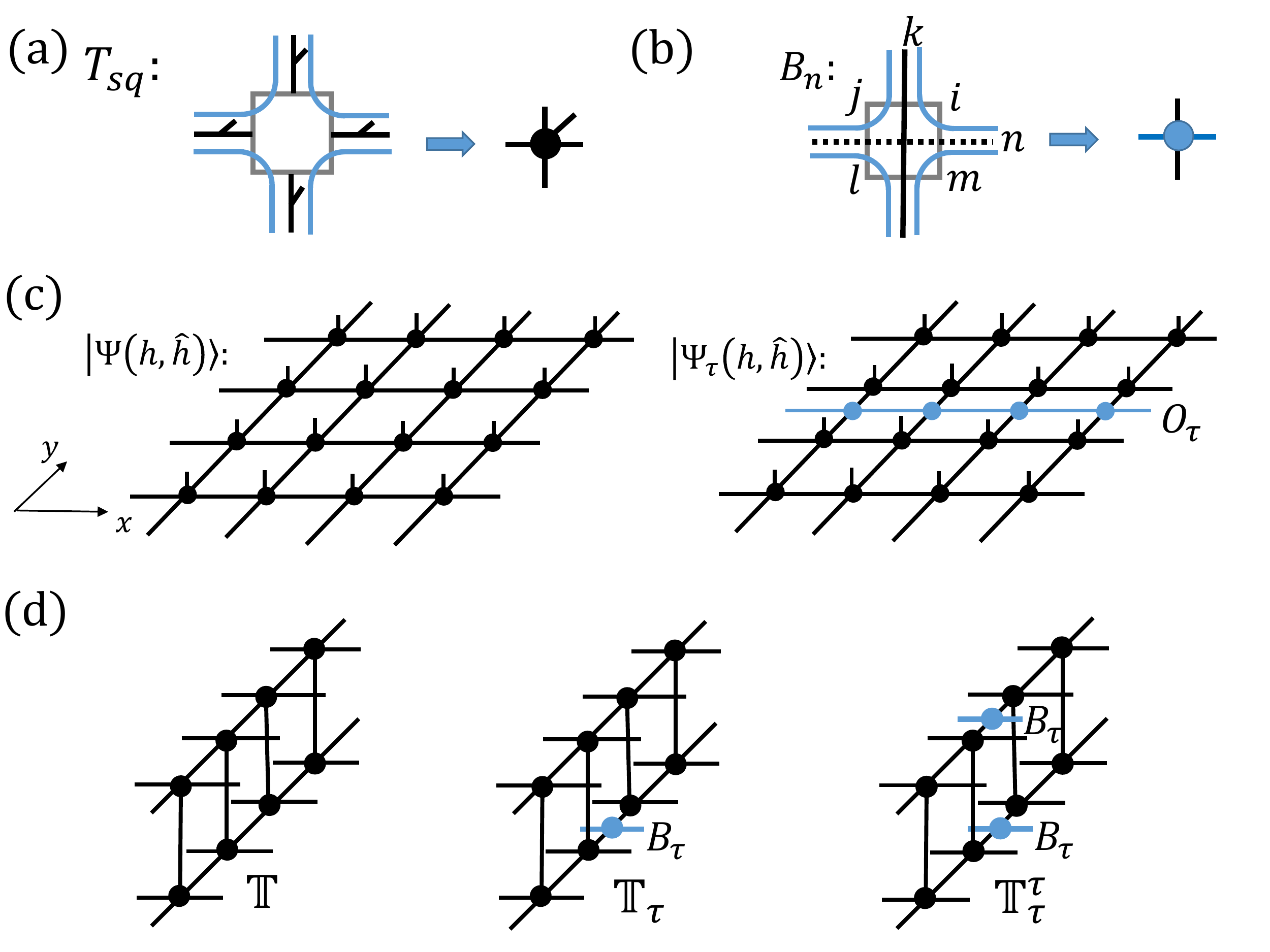}
\caption{(a) A simplified graphic representation for the local tensor $T_{sq}
$ of the deformed wavefunction $|\Psi (h,\hat{h})\rangle $. (b) The local
tensor $B_{n}$ generating the MPO $O_{n}$, where $n=1,\protect\tau $, and
its simplified graphic representation. (c) The tensor networks for
wavefunction $|\Psi (h,\hat{h})\rangle $ and $|\Psi _{\protect\tau }(h,\hat{h})\rangle $ with MPO $O_{%
\protect\tau }$ (blue line) insertion in $x$ direction. (d) Three
inequivalent transfer operators: $\mathbb{T}$ without MPO tensor insertion; $%
\mathbb{T}_{\protect\tau }$ with MPO tensor $B_{\protect\tau }$ insertion in
ket layer and $\mathbb{T}_{\protect\tau }^{\protect\tau }$ with MPO tensor $%
B_{\protect\tau }$ insertion in both layers. }
\label{Transfer_Operator}
\end{figure}

On a torus, the ground state is degenerate. The other
wavefunction denoted by $|\Psi _{\tau }(h,\hat{h})\rangle $ can be obtained
by inserting MPO $O_{\tau }$ into the tensor network, as shown in Fig .\ref%
{Transfer_Operator} (b). The corresponding topological order can be
identified from the transfer operator $\mathbb{T}$ of the wavefunction norm
\begin{equation}
\langle \Psi (h,\hat{h})|\Psi (h,\hat{h})\rangle =\text{Tr}\left( \mathbb{T}%
^{L_{x}}\right) ,
\end{equation}%
where $L_{x}$ is the length of the lattice in the $x$-direction and the
circumference of the transfer operator $\mathbb{T}$ is denoted by $L_{y}$. Since
the wavefunction norm is regarded as two coupled $\phi ^{2}$ Potts models,
the effective dimension of the transfer operator $\mathbb{T}$ is thus given
by $\phi ^{4L_{y}}\times \phi ^{4L_{y}}$ for a large $L_{y}$. The
wavefunction norm thus has the two layers, one is the bra layer and the
other is the ket layer. Inserting the MPO $O_{\tau }$ in either bra layer or
ket layer, or both layers, results in the norms
\begin{eqnarray}
\langle \Psi (h,\hat{h})|\Psi _{\tau }(h,\hat{h})\rangle  &=&\text{Tr}\left[
(\mathbb{T}_{\tau })^{L_{x}}\right] ,  \notag \\
\langle \Psi _{\tau }(h,\hat{h})|\Psi _{\tau }(h,\hat{h})\rangle  &=&\text{Tr%
}\left[ (\mathbb{T}_{\tau }^{\tau })^{L_{x}}\right] .
\end{eqnarray}%
So there are three inequivalent transfer operators $\mathbb{T}$, $\mathbb{T}%
_{\tau }^{\tau }$, and $\mathbb{T}_{\tau }\mathbb{\ }$or $\mathbb{T}^{\tau }$%
, as shown in Fig. \ref{Transfer_Operator} (d), where the subscribe
(superscribe) stands for the MPO tensor $B_{\tau }$ insertion in the ket
(bra) layers.

The anyon sectors of the transfer operators can be classified for the bra
and ket layers individually, and they are denoted by $(\alpha |\beta )$,
where $\alpha ,\beta =1,\tau ,\bar{\tau},b$ are anyon sectors of the bra and
ket, separately. There are in total 16 sectors, in which the topological
sectors $(b|b)$, $(\tau |\tau )$ and $(\bar{\tau}|\bar{\tau})$ of the
transfer operator measuring the confinement of $b$, $\tau $ and $\bar{\tau}$
anyons and sectors $(1|b)$, $(1|\tau )$ and $(1|\bar{\tau})$ measuring the
condensation of $b$, $\tau $ and $\bar{\tau}$ anyons. The dominant
eigenvalues of these sectors determine the property of anyons in the
thermodynamic limit.

Next, we illustrate how to find dominant eigenvalues and the properties of anyons.
By analysis the central idempotent, the layer of a transfer operator
without MPO tensor $B_{\tau }$ insertion contains the anyon sectors $1$ and $b$,
while the layer with MPO tensor $B_{\tau }$ insertion contains the anyon sector $b$,
$\tau $ and $\bar{\tau}$. Therefore, the transfer operator $\mathbb{T}$ includes
three anyon sectors:
\begin{equation}
(1|1)\text{; }(1|b)\text{ \& }(b|1)\text{;\ }(b|b).  \label{sector}
\end{equation}%
$\mathbb{T}_{\tau }^{\tau }$ has the following anyon sectors:
\begin{eqnarray}
&&(b|b)\text{; }(b|\tau )\text{ \& }(b|\bar{\tau})\text{ \& }(\tau |b)\text{
\& }(\left. \bar{\tau}\right\vert b)\text{; }  \notag \\
&&(\tau |\bar{\tau})\text{ \& }(\left. \bar{\tau}\right\vert \tau )\text{, }%
(\tau |\tau )\text{ \& }(\bar{\tau}|\bar{\tau}).
\end{eqnarray}%
And $\mathbb{T}_{\tau }$ contains sectors:
\begin{equation}
(1|b)\text{, }(1|\tau )\text{ \& }(1\left\vert \bar{\tau}\right. )\text{, }%
(b|\tau )\text{ \& }(b|\bar{\tau})\text{, }(b|b).
\end{equation}%
Because of the exchanging $\tau $ and $\bar{\tau}$ under the time reversal
symmetry or the symmetry between bra and ket, some sectors are always
exactly degenerate, and they are denoted by the symbol \textquotedblleft
\&\textquotedblright. Moreover, $\mathbb{T}$ commutes with $O_{\tau
}\otimes O_{\tau }$, the eigenstates of $\mathbb{T}$ are also eigenstates of
$O_{\tau }\otimes O_{\tau }$ with eigenvalue $\phi ^{2}$, $-1$, $-1$, $%
1/\phi ^{2}$ for sectors $(1|1)$, $(1|b)$ \& $(b|1)$, $(b|b)$, separately.
According to the degeneracy of a given level and the associated transfer
operator, we can determine the sectors to which the level belongs.

With the dominant eigenvalues of anyon sectors we concern, we can analysis
the mechanism for quantum phase transition. For
convenience, a quantity is defined $E_{(\alpha |\beta )}=-\log (\frac{%
e_{(\alpha |\beta )}}{e_{(1|1)}})$, where $e_{(\alpha |\beta )}$ is the dominant
eigenvalue of the transfer operator in the sector ${(\alpha |\beta )}$.
So $E_{(1|\beta )}$ measures the condensation of $\beta $ anyon. As the
circumference of the transfer operator $L_{y}\rightarrow \infty $, the
anyons $\beta $ are condensed if $E_{(1|\beta )}\rightarrow 0$ \emph{%
exponentially} with $L_{y}$. On the other hand, $E_{(\alpha |\alpha )}$
measures the confinement of $\alpha $ anyons. As the circumference of the
transfer operator $L_{y}\rightarrow \infty $, the anyons are confined if $%
E_{(\alpha |\alpha )}>0$.

From the viewpoint of anyon condensation, the only
possibility of the Fibonacci topological phase transition is the
condensation of $b$ anyons with the confinement of $\tau $ and $\bar{\tau}$
anyons. We are curious about whether the phase transition along the
self-dual line is beyond the anyon condensation. In analysis below, we make
a reasonable assumption that in the gapped phases the splitting between
degenerate eigenvalues $E_{(\alpha |\beta )}$ below the gap of the transfer
operator spectrum is exponential small with the system size.
\begin{figure}[tbp]
\centering
\includegraphics[width=10cm,trim=0 80 30 50,clip]{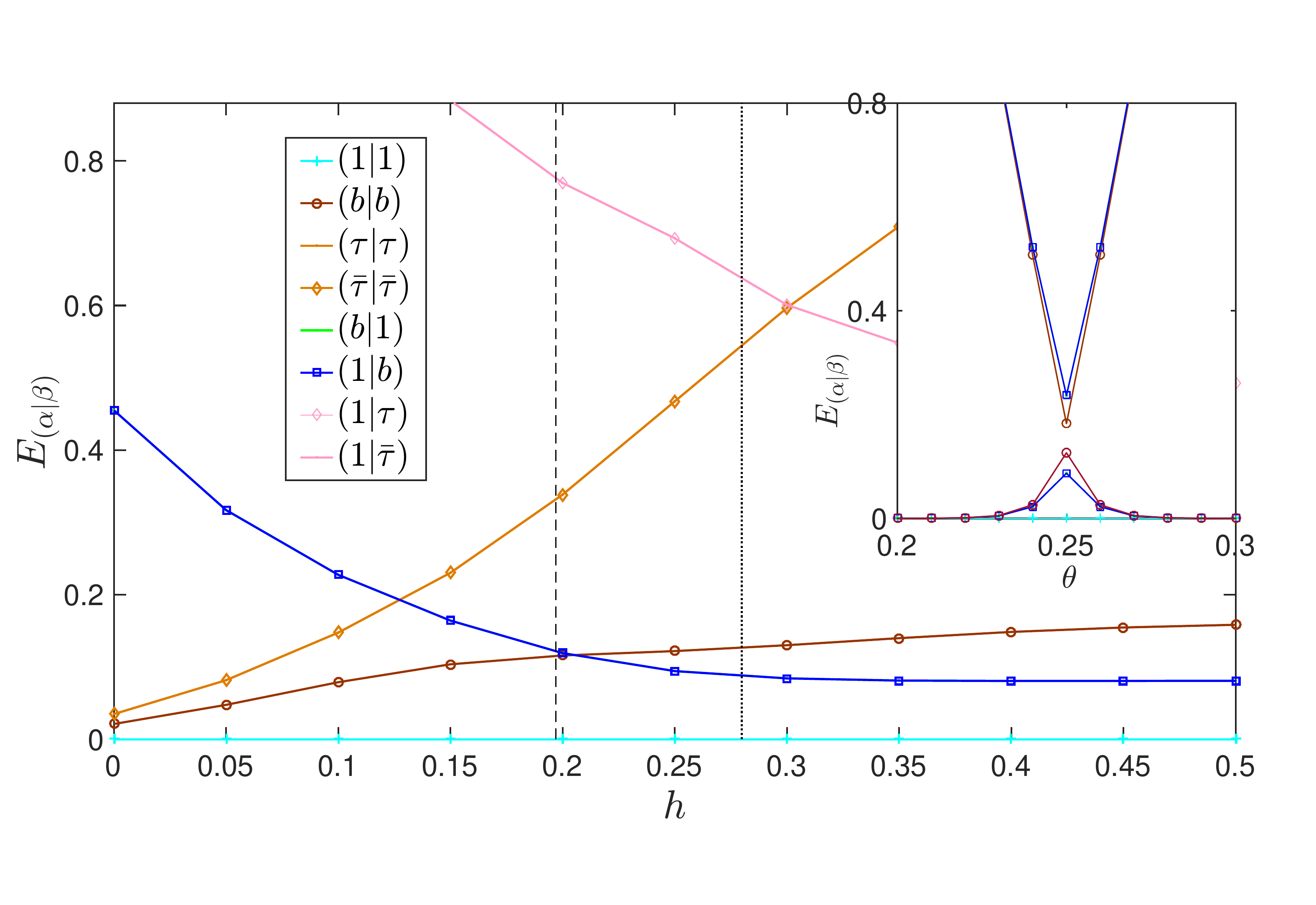}
\caption{The values of $E_{(\alpha |\beta ) }$
for each anyon sectors $( \protect\alpha |\protect\beta ) $
along the self-dual line, obtained from the transfer operator with
circumference $L_{y}=6$. Inset: The spectrum of transfer operator $\mathbb{T}
$ with circumference $L_{y}=6$ along the $\protect\phi ^{4}$-state Potts
lines. The horizontal axis is labelled by angle of polar coordinate so that
the spectrum is symmetric about the self-dual point ($\protect\theta =0.25%
\protect\pi $). }
\label{TO_spectrum}
\end{figure}

In the left panel of Fig. \ref{TO_spectrum} we show the variation of $%
E_{(\alpha |\beta )}$ along the self-dual line. It implies that in the
topological phase, $E_{(b|b)}$, $E_{(\tau |\tau )}$ and $E_{(\bar{\tau}|\bar{%
\tau})}$ approach to zero exponentially, so $b$, $\tau $ and $\bar{\tau}$
anyons are not confined. $E_{(1|b)}$, $E_{(1|\tau )}$ and $E_{(1|\bar{\tau})}
$ are above the spectrum gap, so that the $b$, $\tau $ and $\bar{\tau}$ are
not condensed. On the first-order phase transition line, there is a much
longer correlation length and an approximate conformal invariance.
Thus it is not easy to distinguish the gap and
the degeneracy below the gap on the first-order transition line with such a
small system size. To resolve the problem, we plot the $\phi ^{4}$-state
Potts line in the insert of Fig. \ref{TO_spectrum}, where $\theta =0.25\pi $
corresponds to $h\simeq 0.279$ on the self-dual line. From the inset, the
dominant eigenvalues $E_{(1|b)}$, $E_{(b|1)}$ and $E_{(b|b)}$ and the
subdominant eigenvalues for the three topological sectors are degenerate
with $E_{(1|1)}$ in the thermodynamical limit due to the level crossings at
the first order transition line. The splitting between them is exponentially
small for a finite size system. We can conclude that $E_{(1|b)}\rightarrow 0$
exponentially with $L_{y}\rightarrow \infty $ because it is below the gap.
So along the self-dual line, it still exhibits the $b$ anyon condensation.
In addition, $E_{(b|b)}\rightarrow 0$ exponentially with $L_{y}\rightarrow
\infty $, indicating that $b$ anyons are not confined. Meanwhile,
$E_{(\tau|\tau)}, E_{(\bar{\tau}|\bar{\tau})}>0$ imply the $\tau$ and $\bar{\tau}$
anyons are confined.

\subsection{Finite-size spectrum of conformal field theory}

In the view point of statistical model, the point $D$ in the phase diagram
corresponds to the critical point of two decoupled $\phi ^{2}$-state Potts
models, and we can still calculate the transfer operator spectrum at the
point $D$ to check the validity of our TNS. The single $\phi ^{2}$-state
Potts model at the critical point is described by the second unitary minimal
conformal field theory (CFT) with the central charge $c=7/10$. This CFT is
also the first unitary minimal superconformal model, and it is the only
theory with both Virasoro and super-Virasoro minimal. The scaling dimensions
of primary fields of the CFT are listed in Tab. \ref{7/10 CFT}, and our
numerical results match with such a prediction. The total central charge of
two decoupled models is $7/5$.
\begin{table}[tbp]
\caption{Scaling dimensions of the primary fields in the 2nd unitary minimal
model with the central charge $c=\frac{7}{10}$. The third column is the
results extracted from the transfer operator with circumference $L_{y}=12$
numerically, and the spectrum of the transfer operator is rescaled such that the
scaling dimensions of the first descendents of identity field are 1. The
second column is the Neveu-Schwarz (NS) sector and Ramond (R) sector of the
theory.}%
\begin{tabular}{C{2cm}C{1cm}C{3cm}C{1.2cm}}
\toprule
  scaling dimension& sector & Numerical result&Error \\
\colrule
$0$ & NS &0 &0\\

$3/40$ &R&0.07139148&0.0036\\

$1/5$ & NS &  0.18874674&0.0113\\

 $7/8$ &R &0.84037430&0.0346\\
\botrule
\end{tabular}
\label{7/10 CFT}
\end{table}

At the critical point $A$ on the $\hat{h}$ axis, the critical theory is
described by the seventh unitary minimal CFT with the central charge $c=14/15
$, which belongs to another family of CFT. The corresponding central charge $%
c$ and scaling dimensions $\mathfrak{h}_{r,s}$ of primary fields are given by
\begin{eqnarray}
c &=&1-\frac{6}{m(m+1)},\text{ }m\geqslant 3,  \notag \\
\mathfrak{h}_{r,s} &=&\frac{[(m+1)r-ms]^{2}}{4m(m+1)},\text{ }1\leqslant r<m,%
\text{ }1\leqslant s\leqslant r.
\end{eqnarray}%
Our case corresponds to $m=9$ and the scaling dimensions of the primary
fields are listed in Tab. \ref{14/15CFT}. Notice that only primary fields
with $r=$ odd in the minimal model appear in our microscopic model. The critical theory of the point $B$ has the
same CFT due to the quantum duality.
\begin{table}[tbp]
\caption{Scaling dimensions $\mathfrak{h}_{r,s}+\bar{\mathfrak{h}}_{r,s}$ of
the primary fields in the seventh unitary minimal model with the central charge
$c=\frac{14}{15}$. The third and forth columns are the scaling dimensions
numerically extracted from the transfer operator of the quantum net with
circumference $L_{y}=10$ and their errors. The fifth and sixth columns are
the scaling dimensions numerically extracted from the transfer operator of the
string-net with a circumference $L_{y}=10$ and their errors. The spectrum of
transfer operator is rescaled such that the scaling dimension of the first
descendent of $\mathfrak{h}_{r,s}+\bar{\mathfrak{h}}_{r,s}=0$ is 1. The
absolute errors of data with a star are larger than 10\% compared with exact
scaling dimension. }
\label{14/15CFT}%
\begin{tabular}{C{1cm}C{2cm}C{3cm}C{1.5cm}C{2cm}C{2cm}}
\toprule
 $(r,s)$ &  $\mathfrak{h}_{r,s}+\bar{\mathfrak{h}}_{r,s}$ & Quantum net&Error&String-net&Error  \\
\colrule
 $(1,1)$ & $0$  &$ 0$&$0$&$0$&$0$ \\

 $(3,3)$ & $2/45$ &$0.04973837$& $0.0053$&$0.04369855$&$0.0007$\\

 $(5,5)$ & $2/15$  &$0.14089491$&$0.0076$ &$0.12985540$&$0.0035$ \\

  $(5,5)$ & $2/15$  &$0.14089491$& $0.0076$&$0.12985540$&$0.0035$ \\

 $(7,7)$  & $4/15$ & $0.23521085 $&$0.0315$&$0.25426793$&$0.0124$  \\

 $(2,1)$ & $2/3$   & $0.94338622^*$& $0.2767$&$0.72615165$&$0.0595$  \\

 $(4,3)$ & $14/15$  & $0.94457473$ &$0.0112$&$0.93859015$&$0.0053$  \\

 $(6,5)$  & $56/45$ &$0.95672807^*$&$0.2877$&$1.19624200$&$0.0482$\\

 $(6,5)$  & $56/45$ &$0.95672807^* $ &  $0.2877$&$1.19624200$&$0.0482$ \\

 $(8,7)$ & $8/5$ &$0.98182889^* $ & $0.6182$&$1.44124861$&$0.1588$ \\
\botrule
\end{tabular}
\end{table}

Along the $h$ axis, the effective dimension of the transfer operator $%
\mathbb{T}$ can be reduced from $\phi ^{4L_{y}}\times \phi ^{4L_{y}}$ to $%
(\phi +2)^{L_{y}}\times (\phi +2)^{L_{y}}$, and the number of degrees of
freedom coincides with the $(\phi +2)$-state Potts model. We also calculate
the transfer operator spectrum of the string-net model, which can be
exactly mapped to the $(\phi +2)$-state Potts model. From the comparison
between the transfer operator spectra of the string-net and quantum net, we
suspect that the larger errors for the scaling dimensions of some primary
fields seen Tab. \ref{14/15CFT} may be caused by the fact that the quantum net
model does not exactly correspond to the $(\phi+2)$-Potts model.

Finally, the tri-critical point $C$ in the phase diagram is characterized
by the coset CFT of $\frac{SU(2)_{3}\times
SU(2)_{3}}{SU(2)_{6}}$, which belongs to the series of models with
fractional superconformal symmetry. The central charge $c$ and conformal
dimensions $\mathfrak{h}_{p,q}$ of primary fields for the $\frac{SU(2)_{k}\times
SU(2)_{m}}{SU(2)_{k+m}}$ coset CFT are given by
\begin{eqnarray}
c &=&1-\frac{6m}{(k+2)(k+m+2)}+\frac{2(m-1)}{m+2}, \notag \\
\mathfrak{h}_{p,q} &=&\frac{[(k+m+2)p-(k+2)q]^{2}-k^{2}}{4m(k+2)(k+m+2)}+%
\frac{t (m-t)}{2m(m+2)},
\end{eqnarray}
where $t=(p-q)\bmod m$, $1\leqslant p\leqslant k+1$ and $1\leqslant q\leqslant k+1+m$. The
tricritical point $C$ corresponds to $m=k=3$, and the corresponding
finite-size spectrum has been included in the main text.

\end{widetext}

\end{document}